\documentclass[11pt,hyper]{article}

\usepackage[bulletsep]{collref}
\usepackage{amsmath,epsfig}
\usepackage{amssymb,amsfonts}
\usepackage{epstopdf}
\usepackage{graphicx}
\usepackage{subfigure}
\usepackage{color}
\usepackage{datetime}
\usepackage{bbm}

\newcommand{\bi}{\begin{itemize}}
\newcommand{\ei}{\end{itemize}}

\def\p{\partial}
\def\a{\alpha}
\def\b{\beta}
\def\d{\delta}

\def\l{\lambda}
\def\e{\epsilon}
\def\k{\kappa}
\def\th{\theta}
\def\om{\omega}
\def\s{\sigma}

\def\O{\mathcal{O}}

\def\N{\mathcal{N}}

\def\D{\Delta}
\def\dl{L}

\def\sl2r{SL(2,\mathbb{R})}
\def\sl2{\mathfrak{sl}(2)}

\def\hstar{\hat{\star}}

\def\r{\rightarrow}
\def\half{{\frac12}}

\newcommand{\bea}{\begin{eqnarray}}
\newcommand{\eea}{\end{eqnarray}}
\newcommand{\be}{\begin{equation}}
\newcommand{\ee}{\end{equation}}

\makeatletter \@addtoreset{equation}{section} \makeatother
\renewcommand{\theequation}{\arabic{section}.\arabic{equation}}

\newcommand{\beq}{\begin{equation}}
\newcommand{\eeq}{\end{equation}}
\newcommand\beqa{\begin{eqnarray}}
\newcommand\eeqa{\end{eqnarray}}

\newcommand{\eq}[1]{(\ref{#1})}

\def\cN{{\cal N}}

\def\({\left(}
\def\){\right)}
\def\[{\left[}
\def\]{\right]}

\def\<{\langle}
\def\>{\rangle}

\newcommand{\nn}{\nonumber}

\title{\bf Integrability in dipole-deformed $\mathcal{N}=4$ Super-Yang-Mills}

\author{}

\begin{document}

\begin{flushright}\footnotesize
\texttt{NORDITA-2017-061} \\
\texttt{UUITP-18/17}
\vspace{0.6cm}
\end{flushright}

\renewcommand{\thefootnote}{\fnsymbol{footnote}}
\setcounter{footnote}{0}

\begin{center}
{\Large\textbf{\mathversion{bold} Integrability in dipole-deformed $\mathcal{N}=4$ super Yang-Mills}
\par}

\vspace{0.8cm}

\textrm{Monica~Guica$^{1,2,3}$, Fedor~Levkovich-Maslyuk$^{2}$ and
Konstantin~Zarembo$^{2,3}$}
\vspace{4mm}

\textit{${}^1$Institut de Physique Th\'eorique, CEA Saclay, CNRS, 91191 Gif-sur-Yvette, France}\\
\textit{${}^2$Nordita,  Stockholm University and KTH Royal Institute of Technology,
Roslagstullsbacken 23, SE-106 91 Stockholm, Sweden}\\
\textit{${}^3$Department of Physics and Astronomy, Uppsala University\\
SE-751 08 Uppsala, Sweden}\\
\vspace{0.2cm}
\texttt{fedor.levkovich.maslyuk@su.se, zarembo@nordita.org}


\end{center}
\par\vspace{0.7cm}

\begin{flushright}
{\it \large In memory of Petr Petrovich Kulish}
\end{flushright}

\vspace{2mm}

\begin{center}
\textbf{Abstract} \vspace{5mm}

\begin{minipage}{13cm}

\vskip2mm
 
We study the null dipole deformation of  $\mathcal{N}=4$ super Yang-Mills theory, which is an example of a potentially solvable ``dipole CFT'':  a theory that is non-local along a null direction, has non-relativistic conformal invariance along the remaining ones, and is  holographically dual
to a Schr\"odinger space-time. We initiate the field-theoretical study of the spectrum in this model by
using integrability inherited from the parent theory.
The dipole deformation corresponds to a nondiagonal Drinfeld-Reshetikhin twist in the spin chain picture, which 
renders the traditional Bethe ansatz inapplicable from the very beginning.
We use instead the Baxter equation supplemented
with nontrivial asymptotics,
which gives the full 1-loop spectrum in the $\mathfrak{sl}(2)$ sector. We show that
anomalous dimensions of long gauge theory operators perfectly
match the string theory prediction,
providing a quantitative test of  Schr\"{o}dinger holography.

\end{minipage}

\end{center}

\vspace{0.5cm}

\newpage


\renewcommand{\thefootnote}{\arabic{footnote}}
\setcounter{footnote}{0}

\tableofcontents

\newpage

\section{Introduction}

The $\mathcal{N}=4$ super Yang-Mills (SYM) theory, directly related to string theory via the AdS/CFT duality \cite{Maldacena:1998re} and integrable at the planar level \cite{Beisert:2010jr}, has a wealth of deformations that preserve both of these properties. At the Lagrangian level, these are obtained by replacing the ordinary field product by the $\star$-product \cite{Seiberg:1999vs}; meanwhile, the dual supergravity background is generated by a TsT transformation \cite{Lunin:2005jy}: a sequence of a T-duality and its inverse, interspersed by a shift. The integrable structure is deformed \cite{Beisert:2005if} by the Drinfeld-Reshetikhin twist \cite{Drinfeld:1989st,Reshetikhin:1990ep}. Typical examples of this construction are the $\beta$-deformation \cite{Leigh:1995ep}, which  breaks internal symmetries, and non-commutative theories \cite{Seiberg:1999vs}, with broken space-time symmetries. The dipole deformation  \cite{Bergman:2000cw} occupies an intermediate position between these two cases, with the $\star$-product combining an internal symmetry with a space-time translation.

The dipole deformation in a null direction \cite{Alishahiha:2003ru} is unique in many respects. Being restricted to the light-cone, the null dipole $\star$-product has the mildest possible degree of non-locality (in Euclidean space, the light-cone collapses to a point). While the deformation parameter has  dimensions of length, the resulting theory remains invariant under non-relativistic conformal transformations in the perpendicular directions, and  thus can be viewed as a three-dimensional non-relativistic CFT upon lightlike reduction \cite{Maldacena:2008wh}. 
 Even though the deformation breaks supersymmetry completely, we also
expect the theory to remain exactly solvable, as we will discuss.
 The dual supergravity background, first worked out in \cite{Alishahiha:2003ru} and  revisited in  \cite{Herzog:2008wg,Maldacena:2008wh,Adams:2008wt},  is known as a Schr\"odinger space-time and  is also rather special.  

There are several reasons to study Schr\"{o}dinger spacetimes and their holographic duals, of which  null dipole-deformed $\N=4$ super Yang-Mills  is  
 the only explicit example so far. We list some of them  below.

The best known reason for studying Schr\"{o}dinger  holography is that  Schr\"{o}dinger spacetimes are invariant under non-relativistic conformal symmetries  \cite{Son:2008ye,Balasubramanian:2008dm}, which has promoted them  to  a prototype holographic model for strongly coupled many-body systems with non-relativistic scale invariance. For this application, only the symmetries of the solution are important, and not the detailed Lagrangian, or whether it can be embedded in string theory. The holographic dictionary for such scenarios has been discussed in e.g. \cite{Andrade:2014kba,Guica:2010sw,vanRees:2011fr}. 

Another reason to study Schr\"{o}dinger holography is that it may provide a \emph{tractable}\footnote{
This is because Schr\"{o}dinger holography is a deformation of AdS/CFT with a \emph{tunable} parameter, which reduces to AdS/CFT as this parameter is set to zero \cite{Taylor:2008tg,Guica:2010sw}. This implies that the holographic dictionary can be studied systematically away from the AdS limit using the usual AdS/CFT tools. } example of a \emph{non}-asymptotically AdS, \emph{universal} holographic correspondence. As the example above shows, and as can also be argued using general holography considerations \cite{Guica:2010sw},  the holographic dual of a $d+1$ - dimensional Schr\"{o}dinger spacetime is a $d$ - dimensional field theory, which is non-local along a null direction, while  having non-relativistic conformal invariance along the remaining $d-1$ directions. We will call such a theory a \emph{dipole CFT}, and part of our task is to understand its most general definition and properties\footnote{Dipole CFTs are given by irrelevant, yet finely tuned deformations of a usual CFT, such that they have the same thermal entropy, at large central charge at least, as the original CFT. }. 

An interesting difference between the AdS/CFT and the Schr\"{o}dinger/dipole CFT  correspondences is that in the latter case, the non-relativistic conformal symmetry is \emph{insufficient} to fix the basic holographic dictionary, and additional input from the field theory is needed. This can be already seen at the level of the two- and three-point functions \cite{Fuertes:2009ex,Volovich:2009yh} of low-lying fields.

Let us exemplify this point through the two-point functions. The operators in a dipole CFT are naturally  labeled  by their non-relativistic conformal dimension $\D$ and lightcone momentum $M$. While in a generic non-relativistic CFT, $\D$ and $M$ are \emph{a priori} independent data, in (strongly-coupled) dipole CFTs  dual to  Schr\"{o}dinger spacetimes, bulk locality dictates that $\D$ be a very specific function of the null momentum $M$. For example, for an operator dual to a free bulk scalar of mass $m$, we must have
 
 \be
\label{Nrl-Delta}
\D= \D(M) = \frac{d}{2} + \sqrt{\frac{d^2}{4} +m^2  + \mu^2 M^2}
 \ee
where $\mu$ is a parameter appearing in the Schr\"{o}dinger metric \eqref{schmet}. Thus, unlike in AdS/CFT, where conformal symmetry and large $N$ factorization are sufficient to imply that low-lying operators will be dual to free (local) bulk fields at leading order in $1/N$ \cite{ElShowk:2011ag}, in the Schr\"{o}dinger/dipole CFT holographic correspondence,
 one needs additional input from the field theory to fix the leading holographic dictionary. 
 

In the specific null dipole-deformed $\N=4$ super Yang-Mills theory we study, $\mu^2$ is proportional to the t'Hooft coupling $\l = g^2_{YM} N$, and the formula for the operator dimensions takes a rather similar square-root form.
Thus, we see that the contribution that is not fixed by non-relativistic conformal invariance comes from the renormalization of the scaling dimensions. This is in striking contrast to  $\mathcal{N}=4$ SYM, where all supergravity fields are dual to protected operators. As we will argue in this paper, the additional input that fixes the $M$ dependence of the scaling dimension consists of integrability \emph{and} strong coupling. 



A particularly interesting class of dipole CFTs are the two-dimensional ones, which have been argued to be  related (via DLCQ) \cite{ElShowk:2011cm} to the Kerr/CFT correspondence \cite{Guica:2008mu}, a proposed holographic description for all extremal black holes, including maximally spinning black holes in our own galaxy. An intriguing observation \cite{Detournay:2012dz} is that black hole solutions in asymptotically three-dimensional Schr\"{o}dinger spacetimes (with a set of desirable properties\footnote{We require that the respective solutions reduce to the vacuum $3d$ Schr\"{o}dinger spacetime as the temperature is set to zero, and reduce to the AdS$_3$ black hole solution as the deformation parameter (the analogue of $\mu$) is set to zero. }) have only been found  in consistent truncations of string theory, and these backgrounds are oftentimes integrable. This appears to suggest that mechanisms intrinsic to string theory, and in particular integrability, may play a role in understanding this correspondence. 


To get a handle on 
the non-perturbative structure  of the spectrum and correlation functions of the dipole CFT, we initiate a detailed study of its integrable structure. Integrability for deformations of the $\mathcal{N}=4$ theory, starting with explicit calculations in \cite{Roiban:2003dw,Berenstein:2004ys}, has developed into a full-fledged, general framework in which the $\star$-product is identified with the Drinfeld-Reshetikhin twist of the underlying integrable structure \cite{Beisert:2005if,Ahn:2010ws}
(see  \cite{Zoubos:2010kh} for a short review). Integrable deformations of $\mathcal{N}=4$ SYM have shed new light on simpler, non-supersymmetric theories in which integrability is manifest at the level of planar diagrams \cite{Gurdogan:2015csr,Caetano:2016ydc,Chicherin:2017cns,Gromov:2017cja}, and we believe that integrability can give us a valuable insight into dipole CFTs/Schr\"odinger holography as well. 

In this paper we will focus on the spectral problem in the $\sl2$ sector, which is the simplest subsector nontrivially affected by the deformation. The gauge theory scaling dimensions at one loop are the energy levels of a deformed closed XXX spin chain that is integrable, yet not solvable by the standard Bethe Ansatz due to the absence of a reference state, which in turn is caused by the Jordan cell-type twist. We propose a Baxter equation capturing the spectrum of this spin chain. Its solutions are non-polynomial and have a surprisingly nontrivial asymptotics. From the solutions of this equation we obtain results for the scaling dimensions of  operators with large twist $J$.  These are perfectly matched by the classical string theory predictions that we derive separately. The large $J$ limit corresponds to a Landau-Lifshitz effective theory on the spin chain side, which we also solve and confirm the agreement with the string prediction.

This paper is organized as follows. In section \ref{sptpic}  we discuss in detail the Schr\"odinger background and details of the TsT transformations, as well as construct classical spinning string solutions describing the lowest-lying states in the $\sl2$ sector. In section \ref{FieldPicture}  we review the dipole-deformed SYM theory and describe the framework of Drinfeld twists, together with its implementation for the gauge theory spin chain model. In section \ref{sl2sec}  we discuss in detail the spin chain Hamiltonian, focusing on the $J=2$ case, which we solve using integrability only partially. In section \ref{lll}  we solve the Landau-Lifschitz theory describing the spin chain in the semiclassical large $J$ regime and reproduce the string theory prediction obtained in section \ref{sptpic}. Finally, in section \ref{baxter} we fully develop the integrability techniques and formulate the Baxter equation providing the complete 1-loop spectrum in the $\sl2$ sector. In particular, we reproduce the string theory results in the large $J$ limit. The conclusions are in section \ref{concl}, and the appendices contain technical details.

\section{The spacetime picture \label{sptpic}}

We start this section by  reviewing some basic properties of Schr\"{o}dinger spacetimes and their field theory duals. Then, in subsections \ref{bmn}  and \ref{bmnpoinc},  we present an analogue of the  BMN  spinning string solution for $Sch_5 \times S^5$.

\subsection{Schr\"{o}dinger spacetimes and non-relativistic CFTs}

Schr\"{o}dinger spacetimes were first introduced in \cite{Alishahiha:2003ru} in  the context of string theory  and  later played an important role  in the non-relativistic ``AdS/cold atom'' correspondence \cite{Son:2008ye,Balasubramanian:2008dm}. The metric of a $d+1$ dimensional Schr\"{o}dinger spacetime is given by 

\be
\label{schmet}
ds^2 = - \frac{\mu^2 (dx^+)^2}{z^4} + \frac{2 dx^+ dx^- + dx^i dx_i +dz^2}{z^2} \;, \;\;\;\;\;\;\;\; i = 1, \ldots , d-2
\ee
The solution is usually supported by a massive vector field $A \sim \mu \, z^{-2} dx^+$. One reason this spacetime is interesting is that it  geometrically realises the group of non-relativistic conformal symmetries in $d-1$ dimensions, also known as the  Schr\"{o}dinger group.

The  Schr\"{o}dinger group is simplest to describe through its embedding into  $SO(d,2)$, the $d$-dimensional relativistic conformal group. To define it, one first splits the $d$ coordinates $x^\mu$ into two null coordinates\footnote{For conventions and a full definition, see appendix \ref{schgp}.} $x^\pm$ and  $d-2$ spatial coordinates $x^i$. The Schr\"{o}dinger group consists of those $SO(d,2)$ elements ($P_\mu, K_\mu, D, M_{\mu\nu}$) that commute with $ P_-  = - i \p_-$, the generator of translations along the  $x^-$ lightcone direction:
\be
H = P_+ \;, \;\;\;\;\;\;\; P_i = P_i \;, \;\;\;\;\;\;\; N = P_-
\ee

\be
G_i = M_{i-} \;, \;\;\;\;\;\;\; \mathcal{D} = D + M_{+-}\;, \;\;\;\;\;\;\;M_{ij} = M_{ij}\;, \;\;\;\;\;\;\;C = \half\, K_-
\ee
They consist of translations, both spatial ($P_i$) and along the non-relativistic time coordinate $t=x^+$ ($H$), spatial rotations ($M_{ij}$), 
 Galilean boosts ($G_i$), one special conformal transformation ($C$) and  non-relativistic scale transformations ($\mathcal{D}$), which act as

\be
x^+ \r \l^2 \, x^+ \;, \;\;\;\;\;\; x^i \r \l \, x^i \;, \;\;\;\;\;\; x^- \r x^- \;, \;\;\;\;\;\; z \r \l \, z
\ee
The three generators $\mathcal{D}, H, C$ form an $SL(2,\mathbb{R})$ subgroup of the Schr\"{o}dinger group, which will be important  later. Finally, from the point of view of the non-relativistic symmetries, the generator $N=P_-$ is central, 
since $[N, \;\cdot \;] =0$, and is interpreted either as particle number or as the mass operator. For $N$ to develop a discrete spectrum, the lightlike direction $x^-$ is sometimes compactified \cite{Adams:2008wt,Herzog:2008wg}; however, this implies that the gravity approximation cannot be trusted \cite{Maldacena:2008wh}. We will therefore keep $x^-$ non-compact.

Given this geometric realisation of non-relativistic conformal symmetries, it has been proposed that
 gravitational theories on $Sch_{d+1}$ spacetimes are holographically dual to strongly-coupled non-relativistic CFTs in \emph{two} dimensions less \cite{Son:2008ye,Balasubramanian:2008dm}. However, as the explicit string-theoretical examples indicate, it is more appropriate to think of the holographic duals as being $d$-dimensional field theories that are non-local along the $x^-$ lightcone direction \cite{Herzog:2008wg,Maldacena:2008wh,Adams:2008wt}. These theories, to which we will refer  as \emph{dipole CFT$_d$'s}, are related via dimensional reduction along $x^-$ to $d-1$-dimensional non-relativistic CFT spanning $ t=x^+, x^i$. 

The non-relativistic symmetry structure is very useful in organizing the data of dipole CFTs. A local operator $\O(t,x^i)$ with non-relativistic scaling dimension $\D$ and particle number  $M$ satisfies
\be
[\mathcal{D},\O (0)] = i \D \, \O(0) \;, \;\;\;\;\;\; [N, \O(0)] = iM \, \O(0)
\ee
From the point of view of the dipole CFT, $M$ represents the momentum along the null direction $x^-$ of the corresponding operator, which is naturally labeled by the modes of the Fourier expansion.

Noting that  $C, G_i$ both act as lowering operators of the conformal dimension $\D$, it is possible to define non-relativistic primary operators via \cite{Nishida:2007pj}
\be
[C, \O(0) ] = [G_i, \O(0)] =0
\ee 
Then, as in the case of relativistic CFTs, the operators $H, P_i$ can be used to build an infinite tower of descendants, all of which have the same eigenvalue of the particle number as the corresponding primary. Non-relativistic conformal invariance  fixes  the two-point function of primary operators to  take the form (up to normalization) 
\be
\langle \O_{M}(t,\vec x)\, \O_{M}^\dag (0,0) \rangle = \frac{1}{t^{\D}}\, e^{- i M \frac{|\vec x|^2}{2t}} \label{2pf}
\ee 
and an $n$-point function $(n \geq 3)$ to depend on $n^2-3n +1$ independent cross ratios. These include the usual $n(n-3)/2$ independent cross ratios that one can built from four points, and also a set of $(n-1)(n-2)/2$ non-relativistic three-point invariants \cite{Volovich:2009yh}. This implies in particular that the primary three-point functions can now depend on an arbitrary function of the three-point invariant cross ratio, unlike in relativistic CFTs, where their coordinate dependence is entirely fixed. 

As mentioned in the introduction, in dipole CFTs  dual to  Schr\"{o}dinger spacetimes, bulk locality dictates that $\D$ be a very specific function of the null momentum $M$. The particular functional form depends on the detailed couplings in the supergravity Lagrangian (as can be seen from the specific examples discussed e.g. in \cite{Detournay:2012dz}), but e.g. for an operator dual to a free  massive scalar, 
the scaling dimension takes the form (\ref{Nrl-Delta}) quoted in the introduction.
Similarly, the large $N$ (tree-level) three-point functions computed via holography are given by a very particular function of the non-relativistic cross ratio  \cite{Volovich:2009yh,Fuertes:2009ex}, which coincides with the Fourier transform of the AdS three-point function. 

One very interesting property of non-relativistic CFTs that they share with their relativistic counterparts is the existence of a state-operator map . As shown in  \cite{Nishida:2007pj}, the state $e^{-H} \O(0) | 0 \rangle$ obtained by acting with a primary operator on the Schr\"{o}dinger-invariant vacuum  is an eigenstate of the operator $H_{gl} = H+C$, with an eigenvalue given precisely by the non-relativistic conformal dimension of $\O$, $\D$. Since $C \sim x^2 \p_- $, this corresponds to putting the system in a harmonic potential. 

This map is quite useful in holography, as it allows one to compute the conformal dimension of operators in the dipole CFT at strong coupling as the  energy of particular configurations in the gravitational description, as we will exemplify in section \ref{bmn}. As shown in \cite{Blau:2009gd}, it is possible to find global coordinates in the Schr\"{o}dinger spacetime such that $H_{gl}$ is the generator of translations in global time $T$,  $H_{gl} =-i \p_T$. The global metric reads

\be
ds^2 = - \left( \frac{\mu^2}{Z^4} + 1 \right) dT^2 + \frac{2 dT dV - \vec{X}^2 dT^2 + d\vec{X}^2+dZ^2}{Z^2} \label{glcoord}
\ee
In these coordinates, the particle number generator is $N = - i \p_V$ and
 the harmonic trap potential is reflected in the modification of the boundary metric.

\subsection{Schr\"{o}dinger spacetimes in string theory}

In string theory, Schr\"{o}dinger spacetime solutions can be easily obtained by acting with a solution generating technique known as a TsT transformation/null Melvin twist on a seed $AdS_p \times S^q$ solution \cite{Alishahiha:2003ru,Adams:2008wt,Herzog:2008wg,Maldacena:2008wh}. In this paper, we will be mostly interested in $Sch_5 \times S^5$ solutions of type IIB string theory that are obtained by applying TsT to $AdS_5 \times S^5$, but the analysis goes through in many other dimensions, as well as for space- or time-like dipole deformations \cite{Dasgupta:2000ry,Bergman:2001rw}.  

The TsT transformation is defined as follows. First, one picks a $U(1)$ isometry direction on the $S^5$, corresponding to rotations along an angular direction $\varphi$, and one isometry direction along the AdS boundary - in this case, $x^-$ - and performs the following set of transformations
\bi
\item a T-duality along $\varphi$
\item a shift $x^- \r x^- +  \hat{\mu}\,  \tilde \varphi$, where $\tilde \varphi$ is the T-dual coordinate to $\varphi$
\item a T-duality back along $\tilde \varphi$
\ei
Note that, strictly speaking, one should first apply the above set of 
transformations 
to the D3-brane solution of type IIB supergravity and only then take 
the decoupling limit $\a' \r 0$, while keeping an appropriate combination 
of $\a'$ and $ \mu$ fixed \cite{Alishahiha:2003ru}. 
This more careful definition yields the relation \eqref{relmumuhat} 
between the shift  $\hat \mu$ and the parameter $\mu$ present in 
the supergravity solution.

Let us apply this set of transformations to the AdS$_5 \times S^5$ solution of type IIB supergravity 
\be
ds^2 = \ell^2\, \left(  \frac{2 dx^+ dx^- + dx^i dx_i +dz^2}{z^2}+  (d\psi + P)^2 + ds^2_{\mathbb{C}P^2} \right) \;, \;\;\;\;\; dP = 2 J_{\mathbb{C}P^2} \label{Schback}
\ee
where $\ell$ is the AdS radius in the string frame and we have written the metric on the unit $S^5$ in the form of a $U(1)$ Hopf fibre over $\mathbb{C}P^2$. The solution is  supported by self-dual RR five-form flux 

\be
F_5= 4 \ell^{-1} \left( \om_{AdS_5} + \om_{S^5} \right)
\ee
We pick the sphere isometry direction $\varphi$ to be the $U(1)$ Hopf fibre  $\psi$ and perform the above-described TsT transformation. The solution after TsT reads 

\be
ds^2 =\ell^2\, \left(- \, \frac{\mu^2 (dx^+)^2}{z^4} +  \frac{2 dx^+ dx^- + dx^i dx_i +dz^2}{z^2}+ ds_5^2\right) \;, \;\;\;\;\; \a' B =  \frac{\mu \ell^2 \, dx^+}{z^2} \wedge (d\psi + P) \label{schmetpoinc}
\ee
The parameter $\mu$ is related to the shift $\hat \mu$ as 
\be
\mu = \frac{\ell^2}{\a'} \, \hat \mu = \sqrt{\l} \, \hat \mu=\frac{\sqrt{\lambda }}{2\pi }\,\dl,
 \label{relmumuhat}
\ee
where $\l$ is the t'Hooft coupling and $\dl$ is the parameter that defines the $\star$-product in field theory. 
The $F_5$-form flux is unchanged. Note that TsT has generated an NS-NS B-field, which needs to be taken to infinity in the decoupling limit, with $\a' B$ kept fixed. This background does not preserve any supersymmetry \cite{Maldacena:2008wh}, since all the $\N=4$ SYM supercharges are charged under $\p_\psi$.  However, it is possible to find Schr\"{o}dinger solutions that  preserve up to twelve supercharges \cite{Bobev:2009mw,Donos:2009zf}. 

Another simple (non-supersymmetric) solution   is obtained by performing the TsT along one of the Cartan directions on the $S^5$, which we denote as $\phi_1$. The $S^5$ metric takes the form
\be
ds_{S^5}^2 = d\th^2 + \sin^2\th \, d\phi_1^2 + \cos^2 \th\, d s^2_{S^3}
\ee
Even though the metric at intermediate stages in the TsT transformation is singular, the resulting background is smooth\footnote{This solution is a particular case of backgrounds studied in \cite{Bobev:2009mw,Bobev:2011qx} which are obtained by T-dualizing along an arbitrary Killing vector field on $S^5$. Depending on the choice of the Killing vector the background may or may not preserve supersymmetry.}

\be
ds^2 =\ell^2\, \left(- \, \frac{\mu^2 \sin^2 \th (dx^+)^2}{z^4} +  \frac{2 dx^+ dx^- + dx^i dx_i +dz^2}{z^2}+ ds_{S^5}^2 \right) \label{cartan}
\ee

\be
 \a' B =  \frac{\mu \ell^2 \, dx^+}{z^2} \wedge \sin^2 \th\, d\phi_1
\ee

\medskip

\noindent An interesting fact about  TsT transformations is that they preserve the integrability (if present initially) of the worldsheet $\s$-model for strings propagating in that spacetime. More precisely,
 the worldsheet $\s$-model action for closed strings in these backgrounds can be mapped, by undoing the effects of the TsT transformation, to the action for open strings in $AdS_5 \times S^5$ with particular twisted boundary conditions \cite{Frolov:2005dj,Frolov:2005iq,Alday:2005ww}. For the mixed AdS - sphere TsT at hand, the twisted boundary conditions for open strings in AdS$_5 \times S^5$ read 
\be\label{twist}
 \varphi (\tau ,\sigma +2\pi)-\varphi (\tau ,\sigma )=- 2\pi \hat \mu \, M
\ee

\be
x^-(\tau ,\sigma +2\pi )-x^-(\tau,\sigma  )= 2\pi \hat \mu \, R,
\ee
where $M$ and $R$ are the worldsheet charges associated with translations along $x^-$ and $\varphi$, respectively. 

Integrability of TsT-transformed backgrounds can be established directly, by identifying the string action on such backgrounds with the Yang-Baxter deformation \cite{Klimcik:2002zj,Klimcik:2008eq,Delduc:2013qra,Kawaguchi:2014qwa} of the sigma-model on $AdS_5\times S^5$, as was discussed in detail in \cite{Matsumoto:2014gwa,Matsumoto:2015uja,vanTongeren:2015soa,vanTongeren:2015uha,Osten:2016dvf,Hoare:2016wsk,vanTongeren:2016eeb,Hoare:2016wca,Araujo:2017jkb,Araujo:2017jap},  where the Drinfeld-Reshetikhin twisted symmetry of these models  was also elucidated\footnote{As discussed in \cite{Dlamini:2016aaa} for some examples, the supergravity solution inherits the Hopf algebra structure associated with the Drinfeld-Reshetikhin twist.}. The Schr\"odinger background corresponds to a particular choice of the $r$-matrix defining the Yang-Baxter deformation \cite{Matsumoto:2015uja} and can also be obtained from the coset construction \cite{SchaferNameki:2009xr}.

Thus, one can describe the physics either in terms of closed string theory on the Schr\"{o}dinger background, or in terms of open strings with twisted boundary conditions in AdS. It is a matter of convenience which model to use, as one can be re-expressed through another. The symmetries of the sigma-model, including integrability, are more transparent in the original $AdS_5\times S^5$  formulation, while the TsT-transformed metric is more appropriate for the  low-energy, supergravity analysis, because the low-energy modes of the string are then directly related to the supergravity fields.

\subsection{Spinning BMN-like strings \label{bmn}}

We can exemplify the  discussion above by considering  simple  spinning string  solutions in Schr\"{o}dinger spacetimes, which are the direct analogues of the BMN string \cite{Berenstein:2002jq} in AdS$_5 \times S^5$. The calculation is easiest in the global Schr\"{o}dinger coordinates \eqref{glcoord}, in which there is a simple identification between the energy of the string and the conformal dimension of the dual operator. In the next subsection, we will present the analogous calculation in Poincar\'e coordinates.

Let us consider a pointlike string in global $Sch_5 \times S^5$
\be
\frac{ds^2}{\ell^2} = - \left( \frac{\mu^2}{Z^4} + 1 \right) dT^2 + \frac{2 dT dV - \vec{X}^2 dT^2 + d\vec{X}^2+dZ^2}{Z^2} + (d\psi +P)^2 + ds^2_{\mathbb{C}P^2} \label{glsch5s5}
\ee
rotating with angular frequency $\om$ along the $\psi$ direction\footnote{In principle, we could choose that the string rotate along an arbitrary big circle on the $S^5$; since the B-field does not affect the geodesic solution, the orientation of the big circle does not make any difference.}. 
%
The string traces a geodesic given by
\be
\psi = \om \tau \;, \;\;\;\;\;\; V =  \mu^2 m \tau \;, \;\;\;\;\; T = \k \tau \;, \;\;\;\;\; Z =Z_0= \sqrt{\frac{\k}{m}} \;, \;\;\;\;\; \vec X =0
\ee
The parameters of the solution are related to the conserved charges as 
\be
E_{gl}= \D= \sqrt{\l} \, \k \;, \;\;\;\;\;\; J = \sqrt{\l} \, \om \;, \;\;\;\;\;\; M = \sqrt{\l} \, m 
\label{consch}
\ee
The Virasoro constraint (equivalent to the condition that the geodesic be null) requires that
\be
\k^2 = \om^2 + \mu^2 m^2 \label{virconstr}
\ee
The conformal dimension of the dual operator, which equals the global energy of the spinning string, is thus given by
\beq
\D = \sqrt{J^2 + \mu^2 M^2} =\sqrt{J^2+\lambda \hat{\mu }^2M^2} =  \sqrt{J^2 + \frac{\lambda}{4\pi ^2} \,  \dl^2 M^2}
\label{dbmn}
\eeq
where we have used the relation (\ref{relmumuhat}) between the parameter
$\mu $ and
 the constant shift $\hat{\mu }$ to exhibit
the t'Hooft coupling dependence of this non-relativistic conformal dimension.
 Note that when $\hat{\mu } =0$ or $\l =0$, this reduces to the known dimension of the supersymmetric BMN operator, $\D = J$.

\medskip

As mentioned at the end of the previous section, it is  possible to map this spinning string solution, using the Buscher rules\footnote{Under T-duality along a direction $\vartheta$, we have \cite{Alvarez:1994dn}
\be
\p_\a \vartheta = - \frac{1}{g_{\vartheta\vartheta}} \left(g_{\vartheta \mu} \p_\a x^\mu + i \e_\a{}^\b (B_{\vartheta \mu} \p_\b x^\mu + \p_\b \tilde \vartheta) \right)
\ee
where $\s^\a$ are the worldsheet coordinates, $\vartheta, x^\mu$ are the target space coordinates and $\tilde \vartheta$ is the T-dual coordinate to $\vartheta$.

} to track the transformation of the worldsheet solution as we undo the TsT,
 to  a string solution in $AdS_5 \times S^5$ with non-trivial winding
\be
T = \k \tau \;, \;\;\;\;\; \widetilde{V}=  \mu \om  \s \;, \;\;\;\;\; \tilde \psi = \om \tau - \mu m  \s
\ee
where in these coordinates, the AdS metric is given by \eqref{glsch5s5} with $\mu=0$ and tildes over the coordinates. Thus, we  find explicitly that 

\be
 \widetilde{V} (\tau, \s + 2\pi) - \widetilde{V} (\tau, \s)   = 2\pi \hat{\mu } J \;, \;\;\;\;\;\; \tilde{\psi} (\tau, \s + 2\pi) -\tilde{\psi} (\tau, \s)  = - 2\pi \hat{\mu } M
\ee
which agrees with \eqref{twist}.  The Virasoro constraint will of course yield the same expression \eqref{dbmn} for the non-relativistic energy of the string.

It is instructive to repeat the above calculation for the case of the Schr\"{o}dinger spacetime \eqref{cartan} obtained via the Cartan shift, in which one can consider both strings  that rotate along the shift direction, $\phi_1$, and those that rotate perpendicular to it (along some angle $\phi_2$ inside the $S^3$). The metric is given by \eqref{cartan}, translated to global coordinates. 
 
 For pointlike strings, the geodesic equation requires that either $\theta =0$ (string spinning in the $\phi_2$ plane) or $\theta =\pi/2$ (string spinning in the $\phi_1$ plane). In the first case, the geodesic equation is unchanged from $AdS$, because $\mu \sin \theta =0$ and thus the dual operator has $\Delta = J_2$, just like in the undeformed theory. In the second case, the geodesic  solution is almost identical to the one above, and the dimension of the dual operator is shifted as $\D = \sqrt{J_1^2 + \mu^2 M^2}$. As we will see later, this observation has a direct  counterpart in the deformed $\N=4$ SYM spin chain. 

\subsection{Spinning strings in Poincar\'e coordinates \label{bmnpoinc}}

The BMN solution can also be constructed in  Poincar\'e coordinates,  where it corresponds to a null geodesic that gives a semiclassical description  of the holographic two-point functions \cite{Dobashi:2002ar,Tsuji:2006zn,Janik:2010gc}.

 The starting point is again the string sigma model on the Poincar\'e  Schr\"odinger background (\ref{Schback}). The BMN-like string is a (complex) solution of the equations of motion of the sigma model with the Euclidean worldsheet that depends only $\tau $, has dilatation charge $\Delta $, R-charge $J$, and  momentum conjugate to $x^-$ equal to $M$. The endpoints of the string at $\tau =\pm \infty $ are anchored to the boundary at $\mathbf{x}=\pm \mathbf{X}/2$, $x^+=\pm i T/2$. The geodesic with these properties is given by 
 
\be\label{BMNsolution}
 z=\sqrt{\frac{\kappa T}{2m}}\,\,\frac{1}{\cosh\kappa \tau }\;, \;\;\;\;\;
\mathbf{x}=\frac{\mathbf{X}}{2}\,\tanh\kappa \tau \;, \;\;\;\;\;
x^+=\frac{iT}{2}\,\tanh\kappa \tau 
\ee

\be
x^-=-\frac{i}{2}\left(\frac{\kappa }{m}-\frac{\mathbf{X}^2}{2T}\right)\,\tanh\kappa \tau +i\mu ^2m\tau \;, \;\;\;\;\;
\varphi =i\omega \tau 
\ee
where $\varphi $ is an angle that parametrizes an arbitrary big circle on the sphere.   The parameters of the solution are related to the conserved charges as in \eqref{consch}. The Virasoro constraint requires that $\k$ be determined via \eqref{virconstr},  which then gives the same scaling dimension as a function of the R-charge and the null momentum as in \eqref{dbmn}. 

The BMN solution can be viewed as a result of inserting the vertex operators

\begin{equation}
 V_{\Delta ,M,R}=z^{-\Delta }\,{\rm e}\,^{iMx^-+iR\varphi }
\end{equation}
at $\tau =+\infty $ and $-\infty $ on the cylinder worldsheet. The vertex operators produce  delta-functions in the equations of motion which source the solution \cite{PolyakovStrings2002,Tseytlin:2003ac}. The two-point function of the vertex operators corresponds to the two-point function in the dual CFT:
\begin{equation}
 \left\langle \mathcal{O}\left(\frac{T}{2}\,,\frac{\mathbf{X}}{2}\right)
 \mathcal{O}^\dagger \left(-\frac{T}{2}\,,-\frac{\mathbf{X}}{2}\right)\right\rangle_{CFT}
 =\left\langle V_{\Delta,M,R }V_{\Delta ,-M,-R}\right\rangle.
\end{equation}
At large $\lambda $ and large quantum numbers the worldsheet two-point function can be calculated in the semiclassical approximation.
The action evaluates to zero  so all the contribution comes from the vertex operators:
\begin{equation}
 \left\langle V_{\Delta,M,R }V_{\Delta ,-M,-R}\right\rangle
 \stackrel{{\rm semiclass.}}{=}\lim_{\tau \rightarrow \infty }
 V_{\Delta,M,R }(\tau )V_{\Delta ,-M,-R}(-\tau ).
\end{equation}
Substituting the BMN solution (\ref{BMNsolution}) into the vertex operators we find that the limit makes sense only after the Virasoro constraint (\ref{virconstr}) is imposed. One can show on general grounds that cancellation of divergences at $\tau \rightarrow \infty $ is equivalent to marginality of the vertex operators, so the dispersion relation (\ref{virconstr}) can be alternatively regarded as the on-shell (marginality) condition for the vertex operators that create the BMN string state.  Doing the algebra we get:
\begin{equation}
 \left\langle \mathcal{O}\left(\frac{T}{2}\,,\frac{\mathbf{X}}{2}\right)
 \mathcal{O}^\dagger \left(-\frac{T}{2}\,,-\frac{\mathbf{X}}{2}\right)\right\rangle_{CFT}
 =\frac{\,{\rm const}\,}{T^\Delta }\,\,{\rm e}\,^{-\frac{M\mathbf{X}^2}{2T}}.
\end{equation}
Analytic continuation to real time: $T\rightarrow -it$ gives the answer expected from the non-relativistic conformal Ward identities \eqref{2pf}, where $\D$ corresponds to the non-relativistic conformal dimension of the field.

\begin{figure}[t]
\begin{center}
 \subfigure[]{
   \includegraphics[width=8.2 cm] {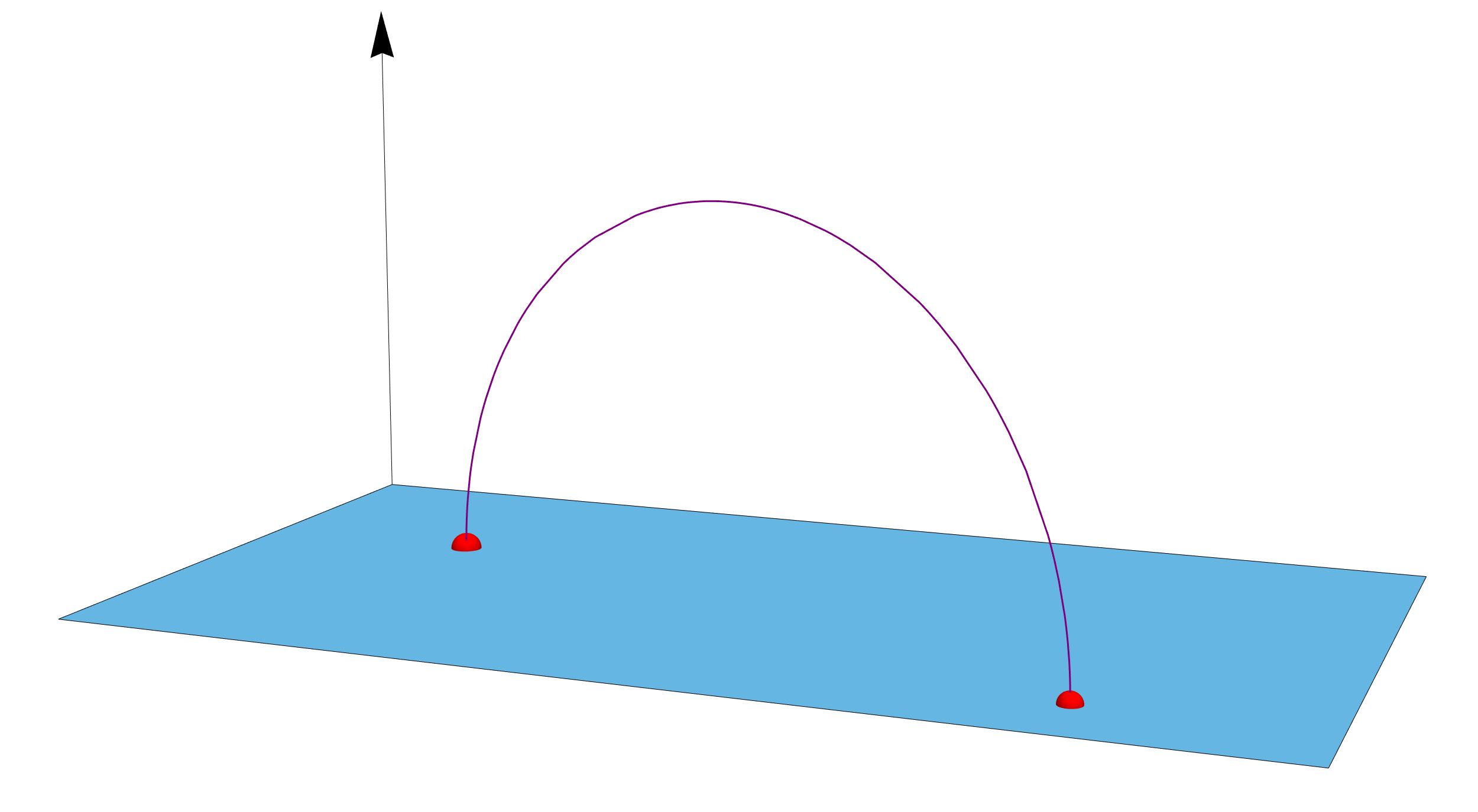}
   \label{fig2:subfig1}
 }
 \subfigure[]{
   \includegraphics[width=8.2 cm] {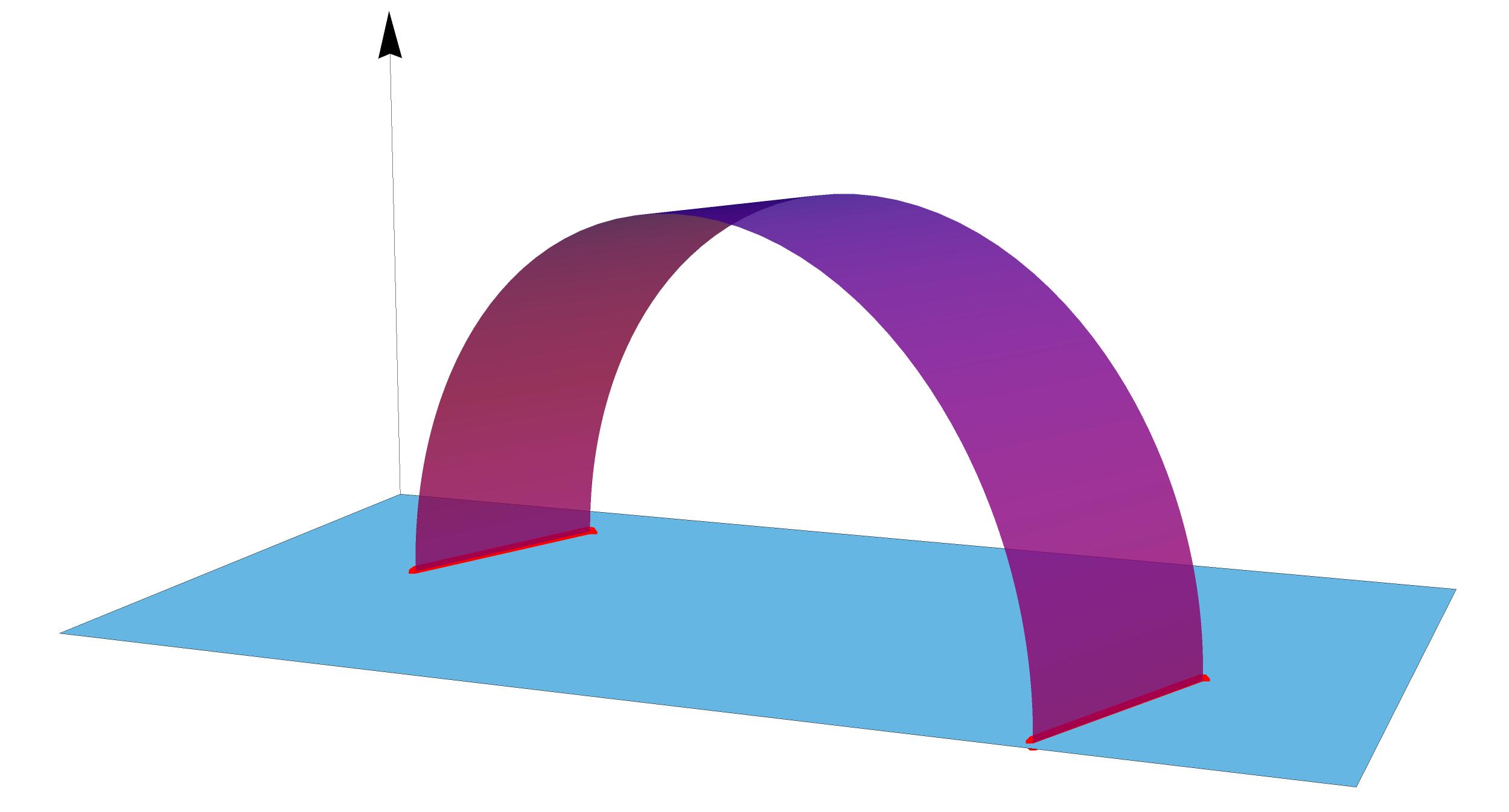}
   \label{fig2:subfig2}
 }
\caption{\label{AdSy}\small The BMN string solution in Poincar\'e coordinates: (a) in the undeformed case, (b) in the presence of the dipole deformation.}
\end{center}
\end{figure}

By undoing the TsT the transformation we find the string solution in Poincar\'e metric of AdS$_5 \times S^5$, given by \eqref{BMNsolution}, together with 
\be\label{mnsol}
{\widetilde{x}}^-=-\frac{i}{2}\left(\frac{\kappa }{m}-\frac{\mathbf{X}^2}{2T}\right)\,\tanh\kappa \tau +\mu \om \s \;, \;\;\;\;\;
\tilde \varphi =i\omega \tau  - m \mu \s
\ee
The twisted boundary conditions on the worldsheet fields are precisely the same as in \eqref{twist}. The undeformed solution with $\mu =0$ connects two points on the boundary as shown in fig.~\ref{fig2:subfig1}, and is naturally associated with point-like, local operators inserted at the ends-points of the string. The solution with $\mu \neq 0$ naturally describes a correlator of two objects that have finite extent along the light-cone, fig.~\ref{fig2:subfig2}. We will see a sharp manifestation of this picture on the field-theory side later when studying one-loop operator mixing in the dual CFT.

\section{The field theory picture}\label{FieldPicture}

\subsection{The dipole deformation in field theory}

The holographic dual of the $Sch_5 \times S^5$ background is obtained by applying a TsT transformation to the theory living on a stack of D3 branes and taking the decoupling limit  \cite{Bergman:2000cw}. As we have seen in the previous section, strings that are charged under the R-symmetry direction become extended along $x^-$ by an amount proportional to their charge. The definition of null dipole-deformed $\N=4$ super Yang-Mills formalizes this intuition. 

One starts by assigning each field in the theory a dipole length, according to its R-charge. We define (half) the dipole length of an operator $\Phi$ with  R-charge $R_\Phi $ to be the null vector 

\be
L_\Phi \equiv \frac{ \dl R_\Phi}{2} \, \p_-
\ee 
where $R_\Phi$ is the R-charge with respect to the sphere $U(1)$ isometry appearing in the TsT. 
If $(J_1,J_2,J_3)$ are the conventional Cartan generators of $SO(6)$, under which $Z_j$ have charges $(1,0,0)$, $(0,1,0)$, and respectively $(0,0,1)$, then for the Hopf fibre twist we are considering we have $R=J_1+J_2+J_3$. 

Next, one replaces the ordinary field products in the  $\mathcal{N}=4$ SYM Lagrangian  by the star product: 

\be
\label{stardefinition}
 (\Phi _1 \star \Phi _2) (x)= \left.\,{\rm e}\,^{\frac{\dl}{2} \left(\partial _{-1}R_2-R_1\partial _{-2}\right)}\Phi _1(x_1)\Phi _2(x_2)\right|_{x_{1,2}=x} = \Phi_1 (x+L_2) \, \Phi_2 (x-L_1) 
\ee
%
%
In order for the star product to be associative, $\Phi_1 \star \Phi_2$ should  be assigned dipole length $L_1 + L_2$, while $L_{\Phi^\dag} = - L_\Phi$ \cite{Dasgupta:2001zu}. In particular, gauge fields, being real, have zero dipole length. The bosonic part of the dipole-deformed SYM Lagrangian reads
\begin{equation}\label{Lagrangian}
 \mathcal{L}_B=
 \frac{2}{g^2}\mathop{\mathrm{tr}}\left(
 -\frac{1}{4}\,F_{\mu \nu }F^{\mu \nu }+\mathcal{D}_\mu Z^\dagger _j\star\mathcal{D}^\mu  Z_j
 -\frac{1}{2}\,[Z^\dagger _j,Z_j]_\star^2
 +[Z^\dagger _j,Z^\dagger _k]_\star[Z_j,Z_k]_\star 
 \right),
\end{equation}
where the covariant derivative too is defined with the star product:
\begin{equation}
 \mathcal{D}_\mu Z=\partial _\mu Z+A_\mu \star Z-Z\star A_\mu .
\end{equation}
As a result, the action is invariant under the star gauge transformations 


\begin{equation} \label{nlgauge}
 \Phi (x)\rightarrow U^{-1}\star\Phi \star U(x)=U^{-1}\left(x+L_\Phi \right)\Phi (x)\, U\left(x-L_\Phi \right).
\end{equation}
instead of the ordinary local gauge transformations. The field with such a transformation law can be pictured as a dipole that extends from $x+L_\Phi $ to $x-L_\Phi $ along the light ray.

In any non-commutative theory, one can construct a locally transforming field variable by a non-local field redefinition, known as the Seiberg-Witten map \cite{Seiberg:1999vs}. In dipole theories the structure of the Seiberg-Witten map is relatively simple. Introducing the gauge connectors,
\begin{equation}
 [x,y]={\rm P}\exp\int_{x}^{y}dv\,A_-(v),
\end{equation}
where integration is along the light ray connecting $x$ and $y$,
the explicit form of the Seiberg-Witten map  reads  \cite{Bergman:2000cw}:
\begin{equation}
 \Phi (x)=[x+L_\Phi ,x]\, \hat{\Phi }(x)\,[x,x-L_\Phi ]=
\parbox{4cm}{\includegraphics[width=4cm]{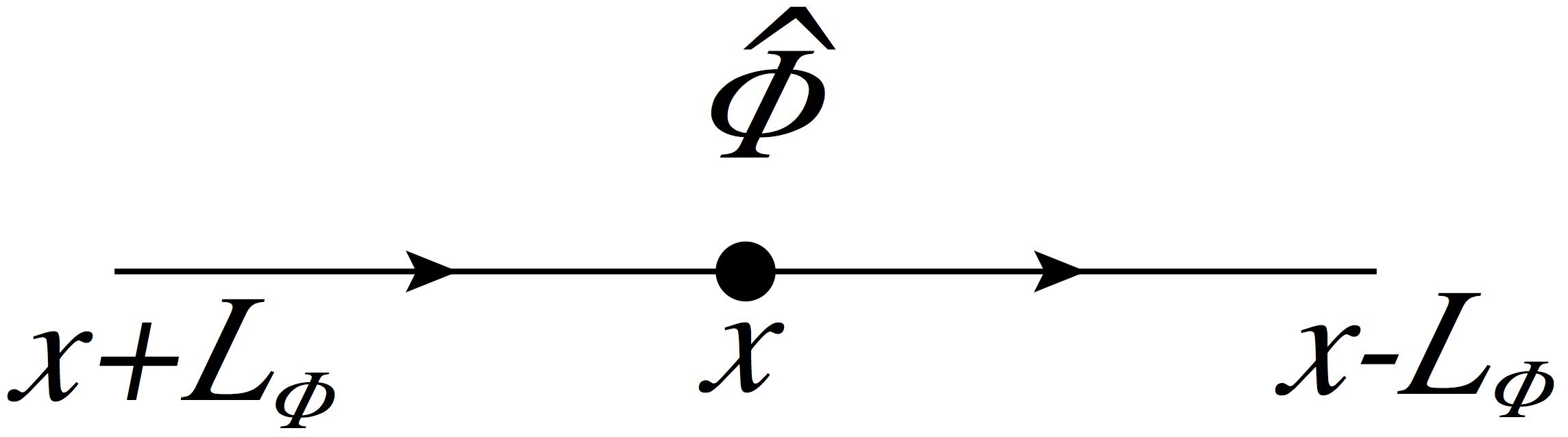}}
\end{equation}
A Wilson line without any insertions can be pictured as the dipole dressing of the unit operator of length $L_{\rm id}$, which we denote by $\mathbbm{1}^{(L_{\rm id})}$: 

\begin{equation}
 \mathbbm{1}^{(L_{\rm id})}=[x+L_{\rm id},x-L_{\rm id}].
\end{equation}
The star product transforms under the Seiberg-Witten map as follows:
\begin{equation}
 \Phi _1\star\Phi _2=
\parbox{4.5cm}{\includegraphics[width=4.5cm]{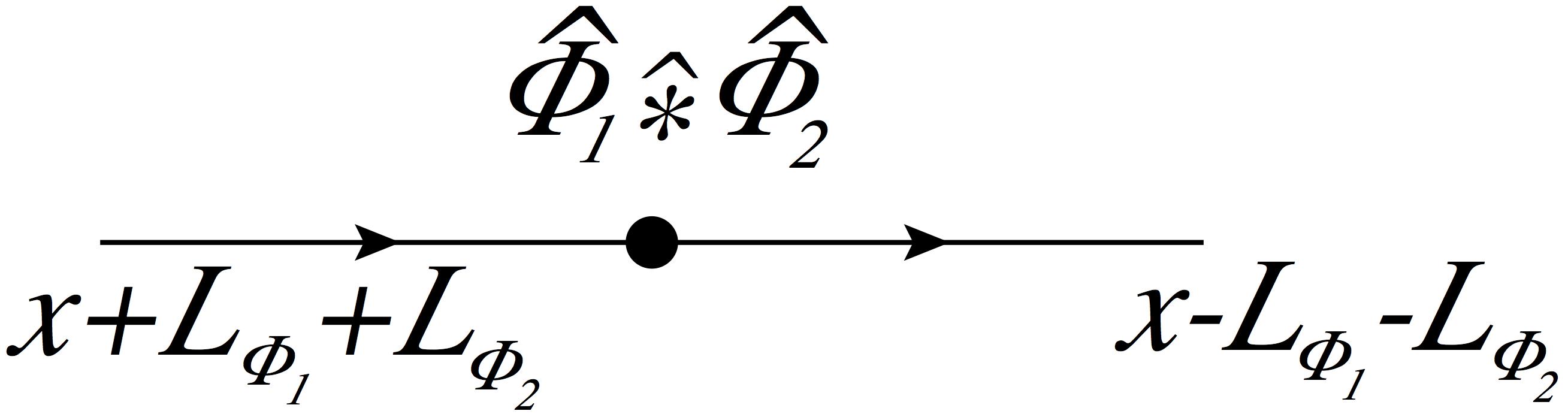}}
\end{equation}
where
\begin{eqnarray} \label{defhatstar}
 \hat{\Phi }_1\hat{\star}\hat{\Phi }_2&=&
\parbox{3cm}{\includegraphics[width=3cm]{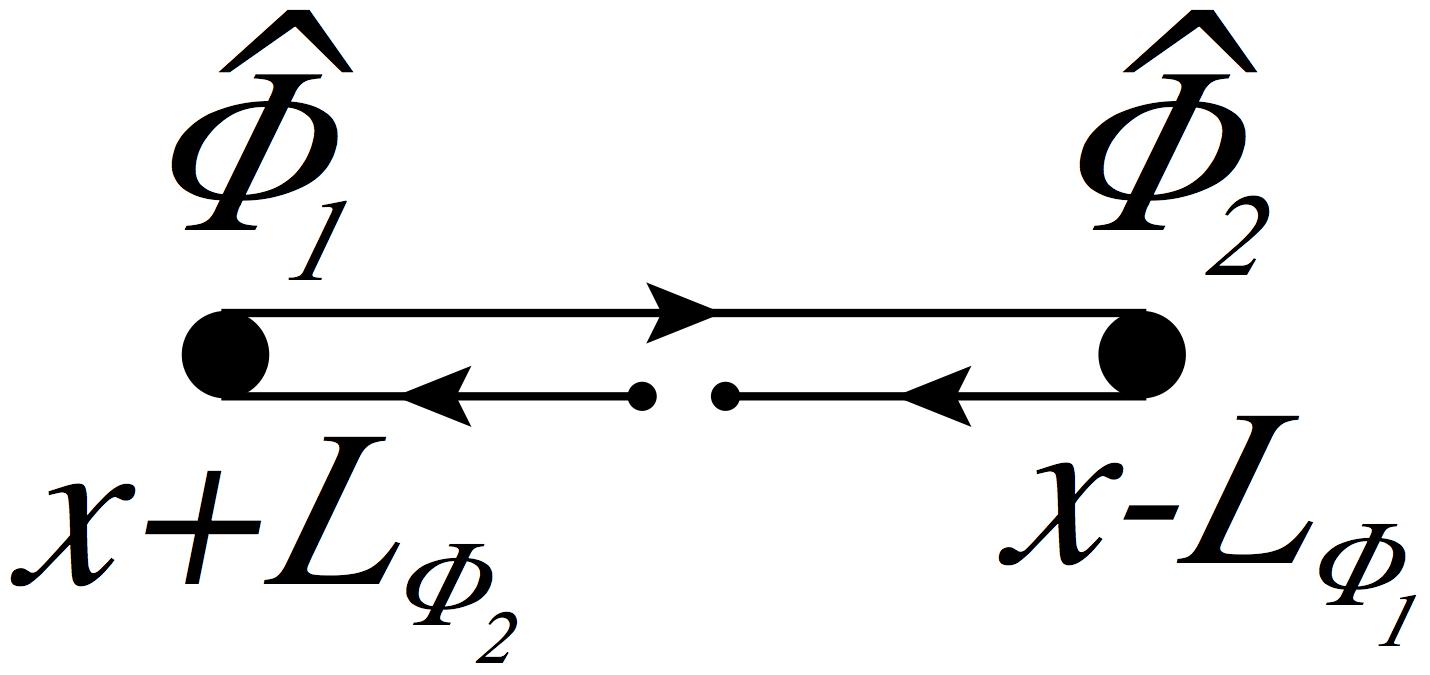}}=
 [x,x+L_{\Phi _2}]\hat{\Phi }(x+L_{\Phi _2})[x+L_{\Phi _2},x-L_{\Phi _1}]
\nonumber \\
&&\times 
\vphantom{\parbox{3cm}{\includegraphics[width=3cm]{Hatstar.jpg}}}
\hat{\Phi }_2(x-L_{\Phi _1})[x-L_{\Phi _1},x]
 =\,{\rm e}\,^{\frac{i\dl}{2} \left(P_{-\,1}R_2-R_1P_{-\,2}\right)}
 \hat{\Phi }_1\hat{\Phi }_2,
\end{eqnarray}
and  $P_-$ is the gauge-covariant momentum operator \cite{Jackiw:1978ar}:
\begin{equation}
 P_\mu \hat{\Phi }=-iD_\mu \hat{\Phi }.
\end{equation}
The covariant derivative remains non-local even after the Seiberg-Witten map:
\begin{equation}
 \mathcal{D}_\mu \Phi =
\parbox{4cm}{\includegraphics[width=4cm]{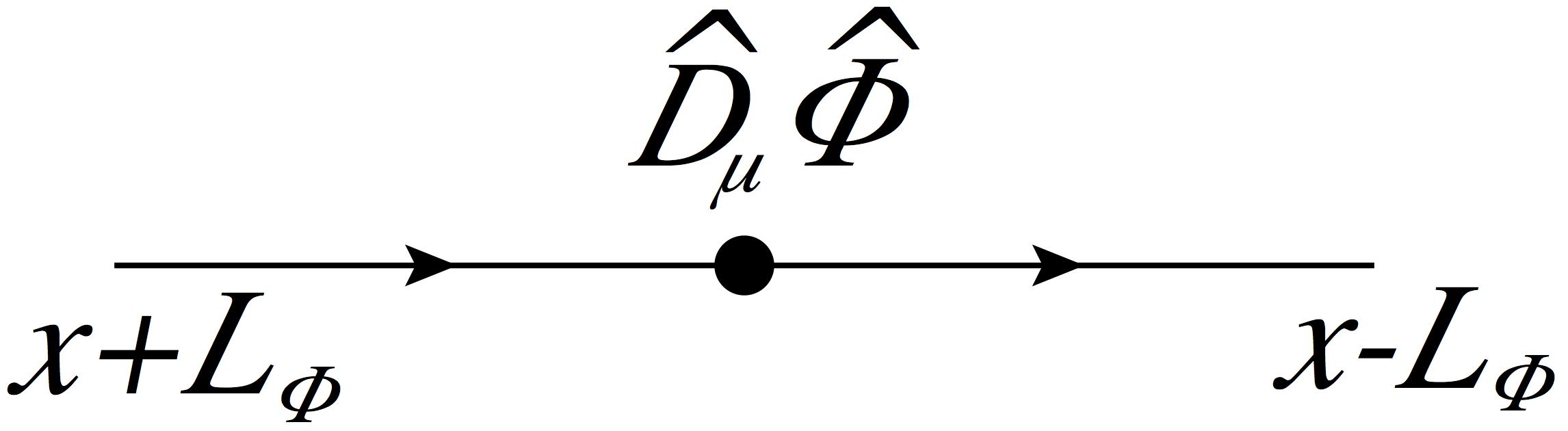}}
\end{equation}
with
\begin{equation} 
 \hat{D}_\mu \hat{\Phi }=D_\mu \hat{\Phi }+\int_{x+L}^{x}dv\,
 \parbox{1.5cm}{\includegraphics[width=1.5cm]{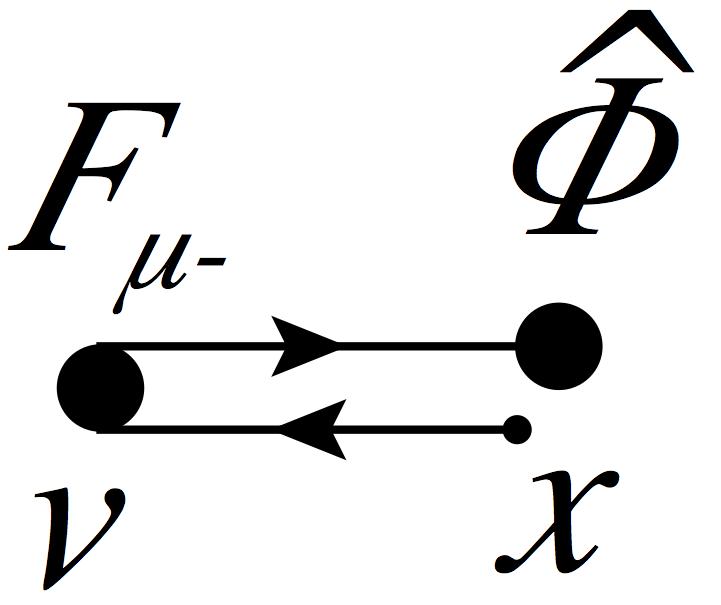}}
 +\int_{x}^{x-L}dv\,
 \parbox{1.5cm}{\includegraphics[width=1.5cm]{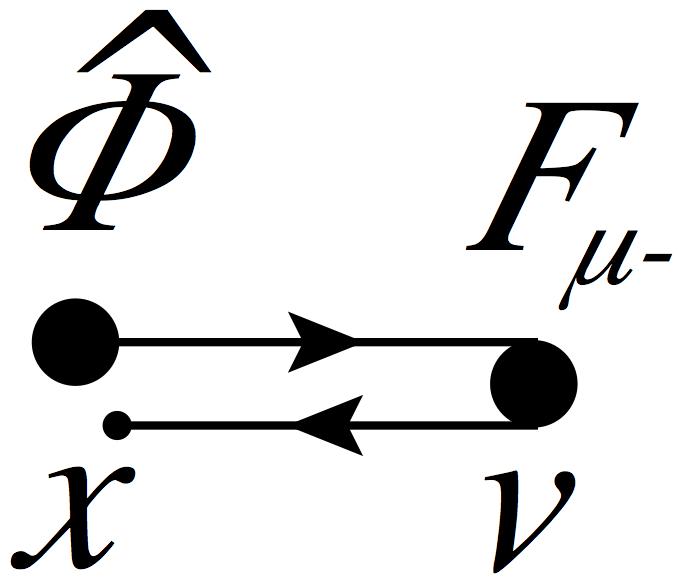}}
\end{equation}
The Seiberg-Witten map trades the (non-local) star product for the manifest non-locality in the interacting part of the Lagrangian:
\begin{equation}\label{Lagrangian-SW}
 \mathcal{L}_B=
 \frac{2}{g^2}\mathop{\mathrm{tr}}\left(
 -\frac{1}{4}\,F_{\mu \nu }F^{\mu \nu }+\hat{D}_\mu \hat{Z}^\dagger _j\hat{D}^\mu  \hat{Z}_j
 -\frac{1}{2}\,[\hat{Z}^\dagger _j,\hat{Z}_j]_{\hat{\star}}^2
 +[\hat{Z}^\dagger _j,\hat{Z}^\dagger _k]_{\hat{\star}} [\hat{Z}_j,\hat{Z}_k]_{\hat{\star}} 
 \right).
\end{equation}
Expanding in $L^\mu$, one obtains an infinite series of higher spin, higher dimension operators \cite{Bergman:2000cw}, all of which are Schr\"{o}dinger invariant in the case of a  null dipole.

The star product, albeit non-local in  coordinate space,  becomes just a phase in the momentum representation. The Feynman diagrams in the dipole-deformed theory thus differ from the ones in  $\mathcal{N}=4$ SYM only by extra momentum-dependent phases in the vertices. These phases can be shifted around using R-charge and momentum conservation, and can be shown to cancel in any planar diagram with no external legs. This is the essence of the planar equivalence theorem: planar vacuum diagrams in the $\star$-deformed theory are the same as in the parent theory, while planar diagrams with   external legs differ by a phase that depends only on the R-charges and momenta of the external lines  \cite{Filk:1996dm}. The planar equivalence theorem  should continue to hold  for gauge-invariant observables after the Seiberg-Witten map, although its justification becomes less obvious due to  extra vertices in the kinetic terms of the Lagrangian (\ref{Lagrangian-SW}).

The planar $\mathcal{N}=4$ SYM and the string sigma model on $AdS_5\times S^5$ are both integrable and are described by a common integrable structure that smoothly interpolates from weak to strong coupling. The planar equivalence theorem as well as compatibility of the integrable structure in the string sigma-model with T-duality guarantees that all $\star$-deformations of $\mathcal{N}=4$ SYM remain uniformly integrable. The deformation can be neatly described as a Drinfeld-Reshetikhin (DR) twist in the boundary conditions of the underlying integrable system \cite{Beisert:2005if,Ahn:2010ws}. The DR twist is the most general abelian transformation of a quantum integrable system that preserves the underlying Yangian algebra \cite{Drinfeld:1989st,Reshetikhin:1990ep}. It is quite remarkable that different $\star$-products in the field theory, possible TsT transformations of $AdS_5\times S^5$ and DR twists of the underlying integrable structure depend on the same set of parameters and can be systematically mapped to  one  another. We will review the construction of gauge-invariant operators in the deformed $\mathcal{N}=4$ SYM, the DR twist of the resulting spin chain, and  the holographic triality between the $\star$-product, TsT transformations and DR twists in the subsequent subsections. 
%

\subsection{Gauge-invariant operators and the map to a spin chain}

\label{sec:sec32}

As we saw, there are two possible, equivalent descriptions of the action of the dipole-deformed theory: either in terms of fields with a local gauge-transformation law, for which the Lagrangian has explicit non-local phases, or in terms of the star product, in terms of which the Lagrangian is identical to that of $\N=4$ SYM (up to $\cdot \r \star$). Both descriptions are of course non-local. Likewise, there are two ways of incorporating the TsT transformation in the string sigma-model: either by quantizing string on the deformed background, or by imposing the twisted boundary conditions on the embedding coordinates of the original sigma-model in $AdS_5\times S^5$. These two pictures are of course equivalent, and it is a matter of convenience which one to use in any particular context. We will call them Model 1 and Model 2. The natural bases of gauge-invariant operators are different in the two pictures, which reflects the difference in their dual string descriptions.

\vskip 0.4cm

\noindent {\bf Model 1}

\vskip1mm

\noindent The elementary building blocks of conformal operators are the field monomials 
\begin{equation}\label{operators-m1}
 \mathcal{O}=\mathop{\mathrm{tr}}\hat{\Phi} _1\ldots \hat{\Phi} _J,
\end{equation}
forming a periodic spin chain of length $J$ due to the cyclicity of the trace.
The adjoint fields at each site are either the scalars $\hat{Z}_j$, $\hat{Z}^\dagger _j$, or fermions, or components of the field strength $F_{\mu \nu }$ or the covariant derivatives thereof, exactly as in $\mathcal{N}=4$ SYM. All elementary fields transform under local gauge transformations, so this picture naturally applies  after the Seiberg-Witten map. The spin chain is periodic and is consequently dual to the closed string on the deformed background.

The spin chain Hamiltonian can be inferred from the planar equivalence theorem, somewhat blurred here by the Seiberg-Witten map, which introduces extra vertices. Most of these additional vertices however vanish in the  light-cone gauge. At any rate, assuming that the planar equivalence holds, planar correlation functions of operators (\ref{operators-m1}) differ from those in the parent $\mathcal{N}=4$ theory by the phases that depend on the quantum numbers of the external lines. The mixing matrix of operators (\ref{operators-m1}) therefore is the same as  $\mathcal{N}=4$ SYM up to a phase transformation. 

For instance, the $\mathcal{N}=4$ one-loop mixing matrix is the Hamiltonian of a nearest-neighbor $PSU(2,2|4)$ spin chain with $J$ sites:
\begin{equation}\label{gamma(1)}
 \Gamma ^{(1)}_{\rm \mathcal{N}=4}=\frac{\lambda }{4\pi ^2}\,\sum_{\ell=1}^{J}h_{\ell,\ell+1}.
\end{equation}
The mixing matrix in the dipole theory will have the same form, where the pairwise spin Hamiltonian  is deformed as follows \cite{Beisert:2005if}:
\begin{equation}\label{tildehFF}
 \tilde{h}_{ab}=F_{ab}h_{ab}F_{ba},
\end{equation}
where
\begin{equation}\label{F-matrix-nulldpl}
F_{ab}=
 \,{\rm e}\,^{\frac{i\dl}{2} \left(P_{-\,a}R_b-R_aP_{-\,b}\right)},
\end{equation}
and $P_{-\,a}$, $R_a$ are the light-cone momentum and the R-charge in the $PSU(2,2|4)$ representation on the $a$-th site of the spin chain.

\vskip 0.4cm\noindent
{\bf Model 2}

\vskip1mm

\noindent One can equally well choose to multiply the fields inside the operator with the star product:

\begin{equation}\label{operators-m2}
 \mathcal{O}=\mathop{\mathrm{tr}}\hat{\Phi} _1\hstar\ldots \hstar\hat{\Phi} _J.
\end{equation}
which is trivially equivalent - using \eqref{defhatstar} -  to considering operators constructed from the dipole fields that obey the non-local gauge transformation law \eqref{nlgauge}

\begin{equation}\label{operators-m2-nc}
 \mathcal{O}=\mathop{\mathrm{tr}}{\Phi} _1\star\ldots \star{\Phi} _J\star\mathbbm{1}^{(-L)},
\end{equation}
where $L$ is the total dipole length carried by the $\Phi _l's$:

\begin{equation}
 L=L_1+\ldots +L_J.
\end{equation}
and the unit field has been added to offset the total dipole length of the product  and make the composite operator gauge-invariant. 

We stress here that the $\star$-product monomials do not constitute a different set of observables compared to the local operators (\ref{operators-m1}), but rather form a different basis in the same space of states. The two bases are related by a similarity transformation, whose explicit form will be given shortly. Any composite operator can be expanded in either one of the two bases - the choice is a matter of convenience.


The planar equivalence theorem applies directly to the set operators defined with the $\star$-product, whose correlation functions are just the same as in $\mathcal{N}=4$ SYM. The spin-chain Hamiltonian in the model 2 therefore is not deformed. What differentiates the deformed theory is the  cyclicity condition of the trace, which is not exactly the same due to non-commutativity of the $\star$-product.  Pulling $\Phi _1$ around the spin chain generates a phase, and the spin chain states as a result are subject to quasi-periodic boundary conditions:
\begin{equation}\label{bc-spinchain}
 \left|s_{J+1}\right\rangle=S\left|s_1\right\rangle,
 \qquad 
 S=\,{\rm e}\,^{i\dl\left(
  P_{-\,1}\mathbf{R}-R_{1}\mathbf{P}_-
 \right)},
\end{equation}
where the boldface letters denote the total charges of the spin chain as a whole. Operationally, twisted boundary conditions mean that the local Hamiltonian on the last link of the chain is defined as
\begin{equation}\label{twist-gen}
 h_{J,J+1}\equiv S^{-1}h_{J,1}S,
\end{equation}
whereas in the untwisted theory $h_{J,J+1}$ is just identified with $h_{J,1}$.

The boundary condition (\ref{bc-spinchain}) is equivalent to the boundary conditions in the string sigma-model, written in  operatorial form. Model 2 thus corresponds to open strings in $AdS_5\times S^5$ with  twisted boundary conditions.

The twist in the boundary conditions can be physically interpreted as a ``magnetic flux'' piercing the spin chain. The same magnetic flux can be described by a constant gauge potential in the local Hamiltonian. The two descriptions are related by an aperiodic gauge transformation\footnote{A nice discussion of twisted boundary conditions in integrable spin chains can be found in \cite{Bazhanov:2010ts}.}. And indeed, the similarity transformation $\Gamma \rightarrow \Omega \,\Gamma \, \Omega^{-1} $ with
\begin{equation}\label{omegaF}
 \Omega =\prod_{l=2}^{J}\prod_{l'=1}^{l-1}F_{l'l},
\end{equation}
when applied to (\ref{gamma(1)}), removes the twist (\ref{twist-gen}) in the boundary conditions and at the same time modifies the local Hamiltonian according to (\ref{tildehFF}), thus mapping model 2 to model 1.

\subsection{The Drinfeld-Reshetikhin twist}

The deformation of the spin chain described above is a particular example of the Drinfeld-Reshetikhin (DR) twist, and as such can be lifted to full integrable structure of the model. Here we briefly review the DR  construction.

 The fundamental building block of an integrable spin chain is the R-matrix \cite{Faddeev:1996iy}, which  acts in the tensor product of two vector spaces $V_a$ and $V_b$, 
 $$R_{ab}(u): \ \ V_a\otimes V_b\rightarrow V_a\otimes V_b,$$ 
 and satisfies the Yang-Baxter equation (YBE):
\begin{equation}
 R_{ab}(u-v)R_{ac}(u)R_{bc}(v)=R_{bc}(v)R_{ac}(u)R_{ab}(u-v).
\end{equation}
The Drinfeld twist \cite{Drinfeld:1989st} is the most general linear transformation of the form
\begin{equation}
 R_{ab}(u)\rightarrow \tilde{R}_{ab}(u)=F_{ab}R_{ab}(u)F_{ab}
\end{equation}
that preserves the YBE. For that to happen, the constant matrix $F_{ab}$ should satisfy two consistency conditions:
\begin{enumerate}
\item $F_{ab}F_{ac}F_{bc}=F_{bc}F_{ac}F_{ab}$
 \item $R_{ab}(u)F_{ca}F_{cb}=F_{cb}F_{ca}R_{ab}(u)$.
\end{enumerate}
In other words, $F_{ab}$ must be a constant solution of the YBE satisfying an intertwining relation with the R-matrix.

The R-matrix acting in the product of an auxiliary space and the quantum space associated with each site of the spin chain can be identified with the Lax operator and used to construct the monodromy matrix
\begin{equation}\label{transfer-matrix}
 T_a(u)=\prod_{\ell=1}^{J}R_{a\ell }(u).
\end{equation}
The monodromy matrix also satisfies the YBE. Its trace over the auxiliary space serves as a generating function for an infinite set of commuting charges \cite{Faddeev:1996iy}.

An R-matrix with isomorphic auxiliary and quantum spaces is regular if at some value of the spectral parameter (taken to be zero) it reduces to permutation operator:
\begin{equation}
 R_{ab}(0)=P_{ab}.
\end{equation}
The spin chain then is translationally invariant, because $U=\mathop{\mathrm{tr}}T(0)$ is the shift operator by one lattice unit \cite{Faddeev:1996iy}, which by construction commutes with all conserved charges of the integrable hierarchy. The first charge in the expansion of $U^{-1}\mathop{\mathrm{tr}}T(u)$ at zero is naturally identified with the Hamiltonian and takes the form (\ref{gamma(1)}) with $h_{ab}$ given by
\begin{equation}\label{localH}
 h_{ab}=\left.\frac{dR_{ab}}{du}\right|_{u=0}P_{ab}.
\end{equation}
Thus, to preserve regularity of the R-matrix, the twist should satisfy an additional constraint:
\begin{enumerate}
\setcounter{enumi}{2}
\item $F_{ab}F_{ba}=1.$
\end{enumerate}
The monodromy matrix $\tilde{T}_a(u)$ constructed from the twisted R-matrices describes a translationally-invariant system once the third axiom is satisfied. The twisted Hamiltonian, derived from (\ref{localH}), then acquires the form (\ref{tildehFF}). The relationship between the twisted spin chain and the untwisted model with modified boundary conditions can be established in full generality. By using just the three axioms above the following relation between the twisted and untwisted transfer matrices can be derived:
\begin{equation}\label{monrelation}
 \tilde{T}_a(u)=\Omega \mathcal{F}_aT_a(u)\mathcal{F}_a\Omega ^{-1},
\end{equation}
where
\begin{equation}\label{Fa-op}
 \mathcal{F}_a=\prod_{\ell=J}^{1}F_{a\ell},
\end{equation}
and 
$\Omega $ is the aperiodic similarity transformation (\ref{omegaF}) that acts only in the quantum space and changes the boundary conditions of the spin chain. The derivation is lengthy but relatively straightforward.

Two broad classes of solutions to the consistency conditions 1-3 exist when the R-matrix possesses symmetries. A non-degenerate linear transformation $K_a$ on the space $V_a$ is a symmetry of the R-matrix if
\begin{equation}\label{commKK}
 K_aK_bR_{ab}(u)=R_{ab}(u)K_aK_b.
\end{equation}
It is straightforward to check that the tensor product 
\begin{equation}\label{ord-twist}
 F_{ab}=K_aK_b^{-1} 
\end{equation}
then satisfies all three axioms. This solution corresponds to the ordinary twist of the R-matrix by the square of $K$:
\begin{equation}
 \tilde{R}_{ab}(u)=K^2_{a}R_{ab}(u)K_b^{-2}.
\end{equation}
Another, perhaps more familiar form of the twist, is obtained with the help of (\ref{monrelation}).
For $F_{ab}$ of the form (\ref{ord-twist}) 
 the operator (\ref{Fa-op}) factorizes:
\begin{equation}
 \mathcal{F}_a=K_a^{J}\mathcal{K}, 
\end{equation}
 where 
\begin{equation}
 \mathcal{K}=\prod_{\ell=1}^{J}K_\ell
\end{equation}
is the symmetry transformation realized on the space of states of the whole spin chain. The transformation law (\ref{commKK}) implies that
\begin{equation}
 K_a\mathcal{K}T_a(u)=T_a(u)K_a\mathcal{K}.
\end{equation}
Multiplying this equation by $K_a^{n-1}$ and taking the trace over the auxiliary space we find that
\begin{equation}
 [\mathop{\mathrm{tr}}K^nT(u),\mathcal{K}]=0
\end{equation}
for any power $n$. Taking trace of the relation (\ref{monrelation}) between the monodromy matrices we find:
\begin{equation}
 \mathop{\mathrm{tr}}\tilde{T}(u)=\Omega \mathcal{K}^2\mathop{\mathrm{tr}}\left(K^{2J}T(u)\right)\Omega ^{-1}.
\end{equation}
The trace of the transfer matrix with the symmetry transformation acting in the auxiliary space,
\begin{equation}\label{tauK}
 \tau_K (u)=\mathop{\mathrm{tr}}K^{2J}T(u),
\end{equation}
can thus be taken as a generating function of conserved charges (transfer matrix) in the twisted spin chain. Since the similarity transformation does not change the spectrum and $\tau (u)$ commutes with $\mathcal{K}$, the eigenvalues of $\mathop{\mathrm{tr}}\tilde{T}(u)$ differ from those of $\tau (u)$ by unimportant phases that only depend on the global charges.

Another class of solutions is associated with a family of commuting charges \cite{Reshetikhin:1990ep}. Let $Q^i$ be such that $[Q^i,Q^j]=0$ and assume that $K=\,{\rm e}\,^{i\omega _iQ^i}$ satisfies (\ref{commKK}) for any $\omega _i$. Then
\begin{equation}
 F_{ab}=\,{\rm e}\,^{\frac{i}{2}\,\gamma _{ij}Q^i_aQ^j_b}
\end{equation}
satisfies axioms 1 and 2. Axiom 3 requires the quadratic form $\gamma $ to be antisymmetric: $\gamma _{ij}=-\gamma _{ji}$. This construction is known as the DR twist. 

In this case $\mathcal{F}_a$ does not factorize, but can be expressed through the total charge of the spin chain: 
\begin{equation}\label{Ftwistmat}
 \mathcal{F}_a=\,{\rm e}\,^{\frac{i}{2}\,\gamma _{ij}Q_a^i\mathbf{Q}^j},
\end{equation}
and the twisted transfer matrix can be identified with 
\begin{equation}
 \tau_\gamma  (u)=\mathop{\mathrm{tr}}\nolimits_a
 \,{\rm e}\,^{\frac{i}{2}\,\gamma _{ij}Q_a^i\mathbf{Q}^j}
 T_a(u)
 \,{\rm e}\,^{\frac{i}{2}\,\gamma _{ij}Q_a^i\mathbf{Q}^j}.
\end{equation}
In particular, the twisted shift operator is given by
\begin{equation}
 U_\gamma \equiv \tau _\gamma (0)=\,{\rm e}\,^{i\gamma _{ij}Q_1^i\mathbf{Q}^j}U,
\end{equation}
 which is equivalent to the twist in the boundary conditions analogous to 
 (\ref{bc-spinchain}).

\subsection{A triality}

Possible DR twists are associated with all possible $\star$-products that one can introduce in field theory:
\begin{equation}
 \Phi _1\star\Phi _2=\,{\rm e}\,^{\frac{i}{2}\,\gamma _{ij}Q_1^iQ_2^j}\Phi _1\Phi _2,
\end{equation}
where $Q_{1,2}^i$ are commuting global charges of $\mathcal{N}=4$ SYM\footnote{Usually $Q^i$ are assumed to be bosonic generators of the symmetry group, but nothing prevents one from including supercharges, provided that the quadratic form $\gamma _{ij}$ is promoted to an appropriately graded supermatrix.}, acting on the fields $\Phi _{1,2}$. Likewise, the most general TsT transformation results in the string boundary conditions
\begin{equation}
 X^M(\sigma +2\pi ,\tau )=\,{\rm e}\,^{i\gamma _{ij}Q^i\mathbf{Q}^j}X^M(\sigma ,\tau ),
\end{equation}
where now $Q^i$ realize the action of commuting isometries on the string embedding coordinates in $AdS_5\times S^5$: $\delta _\epsilon X^M=\epsilon _iQ^iX^M$, and $\mathbf{Q}^i$ are the corresponding Noether charges in the sigma-model. Again, possible TsT transformations are in one-to-one correspondence with possible $\star$-products in the field theory, or DR twists in the spin chain.

\begin{table}[t]
\caption{\label{tablitsa}\small Triality between $\star$-products, TsT transformations, and DR twists. }
\label{dtheories}
\begin{center}
\begin{tabular}[t]{|l|c|c|c|}
\hline
& Field Theory & String Theory & Spin Chain \\
\hline\hline
{\bf Deformation}  & Star product & TsT transformation & DR twist  \\
\hline
{\bf Model 1} & $\mathop{\mathrm{tr}}\hat\Phi _1\ldots \hat \Phi _J$ &
\parbox{5cm}{\begin{center}Closed string on TsT-transformed background\end{center}} & $h_{ab}\rightarrow F_{ab}h_{ab}F_{ba}$ \\
\hline
{\bf Model 2} & $\mathop{\mathrm{tr}}\Phi _1\star\ldots \star\Phi _J$ &
\parbox{5cm}{\begin{center}Open string in $AdS_5\times S^5$  with twisted boundary conditions \end{center}}&
\parbox{5cm}{\begin{center}Spin chain with twisted boundary conditions\end{center}}\\
\hline
\end{tabular}
\end{center}
\end{table}

The triality between the DR twists of the spin chain, possible $\star$-products in field theory, and TsT transformations of the string/supergravity background is summarized in table~\ref{tablitsa}. The data in each case is an anti-symmetric form $\gamma _{ij}$ on the space of commuting charges. The dipole deformation that we are studying here is a specialization of this construction to $\gamma _{P_-R}=-\gamma _{RP_-}=\dl$. 

\subsection{Remarks on the Bethe ansatz}
\label{sec:3last}

Integrability supplies a variety of tools to study the spectrum and the correlation functions of $\mathcal{N}=4$ SYM, and it is usually assumed that the effect of the DR twist on the integrability toolbox is very mild. Let us for definiteness talk about the Bethe ansatz equations. If the charges $Q^i$ that define the DR twist are the same Cartan charges that are used to construct the Bethe ansatz, each Bethe equation will acquire an additional phase that reflects the twist in the boundary conditions experienced by the excitations at the particular nesting level \cite{Beisert:2005if}. More generally, the twist matrix (\ref{Ftwistmat}) has to be first diagonalized by choosing an appropriate Cartan basis. 

It turns out that the dipole deformation affects the integrable structure in a more dramatic way. The light-cone momentum $P_-$ does not belong to the Cartan basis usually chosen to construct the Bethe equations. The twist matrix built from $P_-$ cannot be diagonalized even in principle, because $P_-$  is upper triangular in any highest-weight representation. Let us illustrate this on a simple example.

Consider the subset of operators that are built out of $J$ scalar fields $Z$ and an arbitrary number of light-cone derivatives $D_-$. These operators form a closed $\mathfrak{sl}(2)$ sector under operator mixing, which at one loop is described by a non-compact Heisenberg model, where each spin transforms in the infinite-dimensional spin $-1/2$ representation of $\mathfrak{sl}(2)$ (see e.g. \cite{Derkachov:2002tf, Belitsky:2004sc}). The algebraic Bethe ansatz for this model is constructed from the monodromy matrix in the defining two-dimensional representation of $\mathfrak{sl}(2)$:
\begin{equation}\label{T0sl2}
 T(u)=\begin{pmatrix}
  A(u) & B(u) \\ 
  C(u)  & D(u) \\ 
 \end{pmatrix},
\end{equation}
built according to  (\ref{transfer-matrix})  from the standard $\mathfrak{sl}(2)$ R-matrices
\begin{equation}
 R_{al}(u)=u+i\boldsymbol{\sigma }_a\cdot \mathbf{S}_l.
\end{equation}
The spin variables $\mathbf{S}_l$ can be represented as the differential operators with  $\mathfrak{sl}(2)$ commutation relations defined in  appendix \ref{sec:appsl2}. Thus the entries $A,B,C,D$ are differential operators acting on functions $\psi(z_1,z_2,\dots,z_J)$. The trace of the monodromy matrix $A(u)+D(u)$ commutes with the Hamiltonian and $C(u) $ annihilates the pseudovacuum state given by
\beq
	\left|0\right>=\prod_{j=1}^J \frac{1}{z_j} \ \ .
\eeq
This state is also a common eigenstate of $A(u)$ and $D(u)$. All eigenstates of the transfer matrix ${\rm tr\;} T(u)$ are therefore generated by repeated application of $B(u_i)$ to this reference state, where $\left\{u_i\right\}$ must solve the Bethe equations. 

In the $\mathfrak{sl}(2)$ sector (as in any rank-one sector), the Drinfeld-Reshetikhin twist reduces to an ordinary twist, because the R-charge takes fixed value ($R=1$) at each site, and then the twist matrix (\ref{F-matrix-nulldpl}) factorizes:
\begin{equation}
 F_{ab}=\,{\rm e}\,^{\frac{i }{2}\,\dl \left(P_{-\,a}-P_{-\,b}\right)}\equiv K_aK_b^{-1}.
\end{equation}
In the two-dimensional representation of  $\mathfrak{sl}(2)$, the momentum operator $P_-$  is represented by  $-i\sigma ^-$ .
  The global twist then is given by
\begin{equation}\label{global twist}
K_a^{2J}=\,{\rm e}\,^{\dl J\sigma _a^-}
 =\begin{pmatrix}
 1  & 0 \\ 
 \dl J  & 1 \\ 
 \end{pmatrix},
\end{equation}
and has the form of a Jordan cell. Not possible to diagonalize.

The twisted monodromy matrix of model 2 is related to the monodromy matrix of the undeformed spin chain by eq.~(\ref{tauK}):
\begin{equation}\label{T-mtrx}
 \tilde{T}(u)=\begin{pmatrix}
  A(u) & B(u) \\ 
  C(u) & D(u)  \\ 
 \end{pmatrix}
 K^{2J}
 =
 \begin{pmatrix}
  A(u)+\dl JB(u) & B(u) \\ 
  C(u)+\dl JD(u) & D(u)  \\ 
 \end{pmatrix}
\end{equation}
A simple consistency check of this formula is by dimensional analysis. The $B$ operator creates one extra derivative redistributed along the spin chain and consequently has dimension $1$ in mass units, while $C$ removes one derivative and thus has dimension $-1$. The dipole length $\dl$ has mass dimension $-1$. The diagonal operators $A$ and $D$ are dimensionless. 

The Jordan cell structure of the twist matrix has profound consequences for the spectral equations of the model. In order to apply the usual Bethe ansatz, one should start with a pseudovacuum state, i.e. a common eigenvector of the twisted $A(u)$ and $D(u)$ operators\footnote{Recall that precisely this property is required to prove that eigenvectors of the transfer matrix are obtained by acting on this state with the $B$ operators \cite{Faddeev:1996iy}.}. Typically the pseudovacuum is also annihilated by $C(u)$. However, in the deformed case there are no joint eigenstates of $A(u)$ and $D(u)$, and also no states in the kernel of $C(u)$. This can be checked easily for the spin chain with one site.\footnote{Even though  the state with the wavefunction $\psi(z_1,z_2,\dots,z_J)=1$ is a common eigenstate of the deformed $A$ and $D$ operators, it does not lie in the Hilbert space of the model.} In the absence of a reference state, the Bethe ansatz can be thrown into the paper bin, and from the very beginning one has to deal with more advanced methods based on the Baxter equation, as is typically done in similar context \cite{Faddeev:1994zg}.

These complications should apply of course only to states which carry a nonzero charge $M$ under $P_-$. The $P_-$  operator commutes with the transfer matrix, so they have common eigenstates. In the undeformed model, the spectrum does not depend on $M$ and thus has a high degree of degeneracy. This degeneracy is lifted by  the deformation, so that states with $M\neq 0$ acquire nontrivial energies\footnote{
The change in the spectrum induced by the Jordan cell twist matrix \eq{global twist} is also a special feature of spin chains with an infinite-dimensional Hilbert space. For compact $\mathfrak{su}(2)$ spin chains with a finite-dimensional representation at each site, the spectrum is unaffected by this twist since one can use the $C$ operator to build the states starting from the \textit{dual} pseudovacuum (i.e. the state with all spins up rather than down).}. 

In the next section we will discuss the $\mathfrak{sl}(2)$ sector of the dipole-deformed theory in more detail.

\section{The $\mathfrak{sl}(2)$ sector \label{sl2sec}}

In order to understand the effect of the twist on the gauge theory spectrum, it is useful to work in the smallest closed subsector of gauge theory operators in which its action is non-trivial. This is the $\mathfrak{sl}(2)$  sector,  which is spanned by operators built out of $J$ scalars,
\begin{equation}\label{sl2-op}
 \mathcal{O}=\mathop{\mathrm{tr}}D_-^{S_1}\hat{Z}\ldots D_-^{S_J}\hat{Z},
\end{equation}
where the scalar $\hat Z$  carries R-charge $1$. In the undeformed gauge theory, this sector is closed under renormalization to all orders. On the gravity side, the operators \eq{sl2-op} correspond to strings  moving  on an $AdS_3\times S^1$ subspace of $AdS_5 \times S^5$. 

As we saw in the previous section, the Drinfeld-Reshetikhin twist acts as $e^{i L J  P_- }$  and thus will mix operators with different number of $D_-$ derivatives. In other words, eigenstates of the dilatation operator will not have a definite value of the spin (in contrast to the original theory); instead, they will be eigenstates of $P_-$. Let us stress also that the non-relativistic dilatation generator that we want to diagonalize is a combination of the original dilatation operator and a boost, so it has the form $D+M_{+-}$ (see Appendix \ref{sec:appsl2} for more details). Due to this, at zero coupling all operators \eq{sl2-op} have the same scaling dimension $\Delta$ which is equal to $J$ (the number of scalars in the operator), irrespective of the number of derivatives they contain. The integrable spin chain described above will provide the 1-loop quantum corrections to this classical result, so that
\beq
	\Delta=J+\Delta^{(1)}+O(\lambda^2)\ ,
\eeq
where $\Delta^{(1)}\sim\lambda$ is the 1-loop non-relativistic anomalous dimension.

Let us note also that the string result \eq{dbmn} suggests that for each $J$, already the lowest-lying states, which are protected in the undeformed theory, will acquire anomalous dimensions in the presence of the deformation if they carry momentum $M\neq 0$. We will soon see that this is indeed the case.


In this section we will present a partial solution for the 1-loop spectrum, without employing yet the complete power of integrability. After describing the Hamiltonian, we will focus on the $J=2$ case and discuss the eigenfunctions for $M=0$ or $L=0$. Then, using the fact that the Hamiltonian commutes with the transfer matrix of the algebraic Bethe ansatz, we will find a differential equation for the generic $J=2$ eigenfunctions. This leads to a concise description of the deformed spectrum in terms of known special functions. The solution for any $J$ is postponed to section 6, where it will be obtained via the integrability-based Baxter equation approach.


\subsection{Twist-two operators in the light-ray basis}
\label{J=2}

A convenient basis in the $\mathfrak{sl}(2)$ sector is spanned by the light-ray operators

\begin{equation}\label{confop}
 \mathcal{O}(z_1,\ldots ,z_J)=\mathop{\mathrm{tr}}[0,z_1]\, \hat Z(z_1) \, [z_1,0]\ldots 
 [0,z_J]\, \hat Z(z_J) \, [z_J,0],
\end{equation}
which are related to (\ref{sl2-op}) by  Taylor expansion. The operators mix under renormalization and the conformal operators with definite scaling dimensions are linear combinations of the basic field monomials:
\begin{equation}\label{psiO}
 \mathcal{O}_\psi =\int_{}^{}dz_1\ldots dz_J\,\psi (z_1,\ldots ,z_J)\mathcal{O}(z_1,\ldots ,z_J).
\end{equation}
In the light-ray basis, the operator mixing matrix  is a spin-chain Hamiltonian that is expressed in terms of shifts along the light cone \cite{Balitsky:1987bk}. The dilatation operator acts only on the nearest-neighbor sites:
\begin{equation}
\label{Gham}
 \Gamma =\frac{\lambda }{8\pi ^2}\sum_{l=1}^{J}h_{l,l+1},
\end{equation}
where the local Hamiltonian $h_{ab}$ can be obtained from the dilatation operator 
of the  $\mathcal{N}=4$ SYM in the light-ray representation \cite{Belitsky:2004sc} by applying the Drinfeld-Reshetikhin twist \eq{gamma(1)}, \eq{tildehFF}:
\be
 h_{ab}\, \mathcal{O}(z_a,z_b)=\int_{0}^{1}\frac{du}{u}\,\,
 \bigl[
 2\mathcal{O}(z_a,z_b)-\mathcal{O}(z_a+uz_{ba}+u\dl,z_b)
-\mathcal{O}(z_a,z_b+uz_{ab}-u\dl )
 \bigr]
\ee
where
\begin{equation}
 z_{ab}=z_a-z_b.
\end{equation}
In terms of the wavefunction from (\ref{psiO}) this gives
\begin{eqnarray}
 h_{ab}\,\psi (z_a,z_b)&=&\int_{0}^{1}\frac{du}{u}\,\,
 \left[
 2\psi \left(z_a,z_b\right)-\frac{1}{1-u}\,\psi\left(z_a-\frac{u}{1-u}\left(z_{ba}+\dl\right),z_b\right)
 \right.
\nonumber \\
&&\vphantom{\int_{0}^{1}\frac{du}{u}\,\,}\left.
 -\frac{1}{1-u}\,\psi \left(z_a,z_b-\frac{u}{1-u}\left(z_{ab}-\dl\right)\right)
 \right].
\end{eqnarray}
Since the Hamiltonian commutes with translations, $z_l\rightarrow z_l+c$, the momentum operator $P_-$ and the Hamiltonian can be  simultaneously diagonalized. Explicitly, we have
\beq
	P_-=-i \sum\limits_{j=1}^J \frac{\partial}{\partial z_j}
\eeq
Twist two operators with $J=2$ constitute the simplest example. 
The wavefunctions, which are also eigenstates of $P_-$ with eigenvalue $M$, then have the form
\begin{equation}
\label{psichi}
 \psi (z_1,z_2)=\chi (z)\,{\rm e}\,^{\frac{i}{2}M\left(z_1+z_2\right)}\;, \;\;\;\;\;\; z = z_2-z_1
\end{equation}
The Schr\"odinger equation for the relative wavefunction is 
\begin{eqnarray}
\label{schrc}
 &&\int_{0}^{1}\frac{du}{u}\,\,\left[
 2\chi (z)-\frac{1}{1-u}\,\cos\frac{Mu\left(z+\dl\right)}{2(1-u)}\,\,
 \chi \left(\frac{z+u\dl}{1-u}\right)-
 \right.
\nonumber \\
&&\hspace{3.4cm}\vphantom{\int_{0}^{1}\frac{du}{u}\,\,}\left.
 -\frac{1}{1-u}\,\cos\frac{Mu\left(z-\dl\right)}{2(1-u)}\,\,
 \chi \left(\frac{z-u\dl}{1-u}\right)
 \right]=E\,\chi (z).
\end{eqnarray}
The eigenvalue is related to the 1-loop anomalous dimension for $J=2$ as
\begin{equation}
 \Delta^{(1)} =\frac{\lambda }{4\pi ^2}\,E.
\end{equation}
Below we will describe how to solve this eigenvalue equation, starting with the simpler cases when either $M$ or $\dl$ is zero.

\subsection{Explicit solutions at zero momentum}

In order to solve the eigenvalue equation for the Hamiltonian it is convenient to switch to Fourier space. In terms of the  Fourier transformed wavefunction $\chi(p)$
\begin{equation}
 \chi (z)=\int_{-\infty }^{+\infty }\frac{dp}{2\pi }\,\,\,{\rm e}\,^{ipz}\chi (p),
\end{equation}
the Schr\"odinger equation takes the form
\begin{eqnarray}\label{sch2}
 &&\int_{0}^{1}\frac{du}{u}\,\,
 \left[2\chi \left(p\right)-\cos \dl u\left(p+\frac{M}{2}\right)\,\chi \left((1-u)p-\frac{M}{2}\,u\right)
 \right.
\nonumber \\
&& \hspace{2cm}\vphantom{\int_{0}^{1}\frac{du}{u}\,\,}\left.
 -\cos \dl u\left(p-\frac{M}{2}\right)\,\chi \left((1-u)p+\frac{M}{2}\,u\right)\right]
 =E\chi \left(p\right).
\end{eqnarray}
In the undeformed theory with $\dl=0$ the solutions to the latter equation are polynomials:
\begin{equation}
 \chi _S(p)=p^S+a_{S-1}p^{S-1}+\ldots +a_0,
\end{equation}
corresponding to the operators
\begin{equation}
 \mathcal{O}_S=\mathop{\mathrm{tr}}ZD_-^SZ+{\rm descendants}.
\end{equation}
If $M=0$ they reduce simply to
\beq
\label{chips}
	\chi _S(p)=p^S \ .
\eeq
The eigenvalue can then be read off from the action of (\ref{sch2}) on the highest power:
\begin{equation}
 E_S=2\int_{0}^{1}\frac{du}{u}\,\,
 \left[1-\left(1-u\right)^S\right]=2h(S),
\end{equation}
where $h(S)$ are the harmonic numbers. The spectrum does not depend on $M$, as expected in the undeformed theory.

In general, we expect that the spectrum should depend only on $M\dl$ and thus should remain the same for $M=0$ when the deformation is switched on, i.e. when $\dl$ becomes nonzero. For $M=0$ the equation \eqref{sch2} takes the form
\begin{eqnarray}
\label{chiM0}
 &&2\int_{0}^{1}\frac{du}{u}\,\,
 \left[\chi \left(p\right)-\cos (up\dl)\,\chi \left((1-u)p\right)
 \right]
 =E\chi \left(p\right).
\end{eqnarray}
To see that the spectrum remains the same, we can expand the  cosine for small $\dl$, so as to obtain the original Hamiltonian plus a perturbation. Writing the perturbation as an infinite matrix in the original eigenbasis (which is given by $p^n$ with $n=0,1,2,\dots$) we see that it is lower triangular with zeros on the diagonal, to all orders in $\dl$. Therefore, the spectrum of the new Hamiltonian is indeed the same as before.
%

The explicit eigenfunctions are given by 
%
%
%
spherical Bessel functions\footnote{This can be easily  checked by showing that  both sides of \eqref{chiM0} are annihilated by the  differential operator $D = p^2 \p_p^2 + 2 p \p_p + p^2 - S(S+1)$.
In practice, we first found the solutions for the first few $S$ by perturbatively expanding in $\dl$ around the $\dl=0$ wavefunctions \eq{chips} as $\chi_S(p)={\rm const\;}(p^S+O(\dl))$.
}
\beq
\label{chism}
	\chi_S(p)=\frac{J_{S+1/2}(p)}{\sqrt{p}}
\eeq
In particular, the deformed ground state corresponds to $S=0$, which gives
\beq
	\chi_{0}(p)={\rm const} \times \;\frac{\sin(\dl p)}{\dl p}
\eeq
It is also interesting to discuss these wavefunctions in coordinate, rather than momentum space.  In the undeformed theory with $M=0, \dl=0$, the solutions \eq{chips} lead to
\beq
	\chi_S(z)={\rm const} \;\delta^{(S)}(z)
\eeq
i.e. they give derivatives of the delta-function concentrated at $z=0$. When we switch on the deformation parameter $\dl$, the delta-function becomes smeared out and is replaced by a step function, as one can see from the Fourier transform of the eigenfunctions 
 \eq{chism},  which reads 
\beq
\label{chiL}
	\chi_S(z)=P_S\(\frac{z}{\dl}\)\[\Theta\(1-\frac{z}{\dl}\)-\Theta\(-1-\frac{z}{\dl}\)\]
\eeq
where $\Theta$ is the Heaviside step function and $P_S$ are Legendre polynomials\footnote{Alternatively, one could start from $1/z^S$ as the $M=0$ eigenstates in the undeformed theory. However we will follow the approach discussed here, as it generalizes concisely to the case of generic $M$ and $\dl$.}.

At this point one can already compute the energy as an expansion in $M$ by developing usual perturbation theory in the basis of functions \eq{chiL}. However in the next section we will see that one can access the wavefunctions and energy at \textit{any} $M$ and express them in terms of known special functions.


\subsection{Algebraic Bethe ansatz and explicit solution}

While the Hamiltonian is an integral operator, it commutes with the trace of the monodromy matrix \eq{T-mtrx}, which can be realized as a differential operator. The resulting differential equation is easier to study than the integral eigenvalue equation \eq{chism} and in this section we will explore it in detail.

As discussed in section \ref{sec:sec32}, one can work within either model 1 (where the R-matrix is changed) or model 2 (where the boundary conditions are twisted). Although we wrote the Hamiltonian \eq{chism} in model 1, it is more convenient to study the transfer matrix in model 2. The eigenfunctions in the two models are related by a simple transformation \eq{omegaF}, which   for $J=2$ is given by
\beq
\label{psi12}
	\psi_1(z_1,z_2)=\psi_2(z_1,z_2+\dl)
\eeq
%
The monodromy matrix in the deformed case is given by \eq{T-mtrx},
\begin{equation}
\label{Ttw}
 T(u)=\begin{pmatrix}
  A(u)+\dl JB(u) & B(u) \\ 
  C(u)+\dl JD(u) & D(u)  \\ 
 \end{pmatrix}
\end{equation}
where $A,B,C,D$ are differential operators as explained in section \ref{sec:3last}, acting on functions of $z_1$ and $z_2$ in the $J=2$ case.
The transfer matrix which we want to diagonalize reads
\beq
\label{TAB}
	{\rm tr}\; T(u)=A(u)+D(u)+\dl JB(u) \ .
\eeq	
As we are considering states with a well-defined value $M$ of the momentum, the wavefunctions have the form 
\begin{equation}
\label{psif}
		\psi_2(z_1,z_2)=f(z)\,{\rm e}\,^{\frac{i}{2}\,M(z_1+z_2)} \;, \;\;\;\;\; z=z_2-z_1
\end{equation}
Acting on such a state with the transfer matrix \eq{TAB}, we obtain the differential equation 
\begin{eqnarray}
\label{fzML}
&&z (-z+2 \dl) f''(z)+2 (\dl- z) f'(z)
\vphantom{\frac{1}{2}}
\nonumber \\&& \hspace{2cm}+\frac{1}{2}\, f(z) \left[\dl M (M z+4
   u)-\frac{1}{2}\, M^2 z^2+4 u^2-1\right]=t(u)f(z)\ ,
\end{eqnarray}
where $t(u)$ is the eigenvalue of ${\rm tr}\; T(u)$. From the terms linear and quadratic in $u$ in this equation we read off
\beq
	t(u)=2u^2+2\dl Mu+t_0 \ ,
\eeq
where $t_0$ is a constant. The $u^0$ term gives a nontrivial differential equation which after the change of variables 
\beq
\label{fG}
	f(z)=G(z-\dl)
\eeq
takes the form
\beq
\label{eqsm}
	(\dl^2-z^2)G''(z)-2zG'(z)+M^2(\dl^2-z^2)G(z)=(t_0+1/2)G(z) \ .
\eeq
Notice that combining \eq{psi12}, \eq{psif} and \eq{fG}, we find that the wavefunction in model 1 is related to $G(z)$ according to
\beq
	\psi_1(z_1,z_2)=G(z_2-z_1)e^{\frac{i}{2}M(z_1+z_2)} \ .
\eeq
Comparing this with \eq{psichi}, we see that $G(z)$ should be identified with the eigenfunction $\chi(z)$ of the Hamiltonian \eq{schrc}. One can verify directly that the Hamiltonian commutes with the differential operator on the l.h.s. of \eq{eqsm}, so they have common eigenfunctions\footnote{To prove that they commute we integrate by parts repeatedly, and the boundary terms which appear will vanish provided that $G(z)$ decays fast enough on the real line for $z\to\pm\infty$.}.

The differential equation \eq{eqsm} defines an eigenvalue problem, and we expect that the solution $G(z)$ should be fixed by requiring it to be regular at $z=\pm \dl$. 
As a check, when $M=0$ the regular solutions of \eq{eqsm} are Legendre polynomials,
\beq
	G(z)=P_S(z/\dl)\ ,
\eeq
precisely reproducing the Legendre polynomials appearing in the explicit wavefunctions $\chi_S(z)$ we found in the previous section (see \eq{chiL}). There is a subtlety though, as those wavefunctions also include a combination of step functions which restricts them to the domain $-\dl<z<\dl$. We expect that, similarly to this case, the Hamiltonian eigenfunctions in general can be reconstructed from $G(z)$ as
\beq
	\chi(z)=G(z)\[\Theta\(1-\frac{z}{\dl}\)-\Theta\(-1-\frac{z}{\dl}\)\] \ .
\eeq
The final extent of the wavefunction along the light-cone is not surprising, as we are dealing with a non-local field theory. The fact that the eigenstates have exactly the dipole length (the wavefunction is strictly zero beyond the interval $(-L,L)$) fits nicely with the string-theory picture of the BMN operators displayed in fig.~\ref{fig2:subfig2}, even though here we are dealing with a different range of parameters (weak coupling, short operators).

\begin{figure}[t]
\begin{center}
 \centerline{\includegraphics[width=14cm]{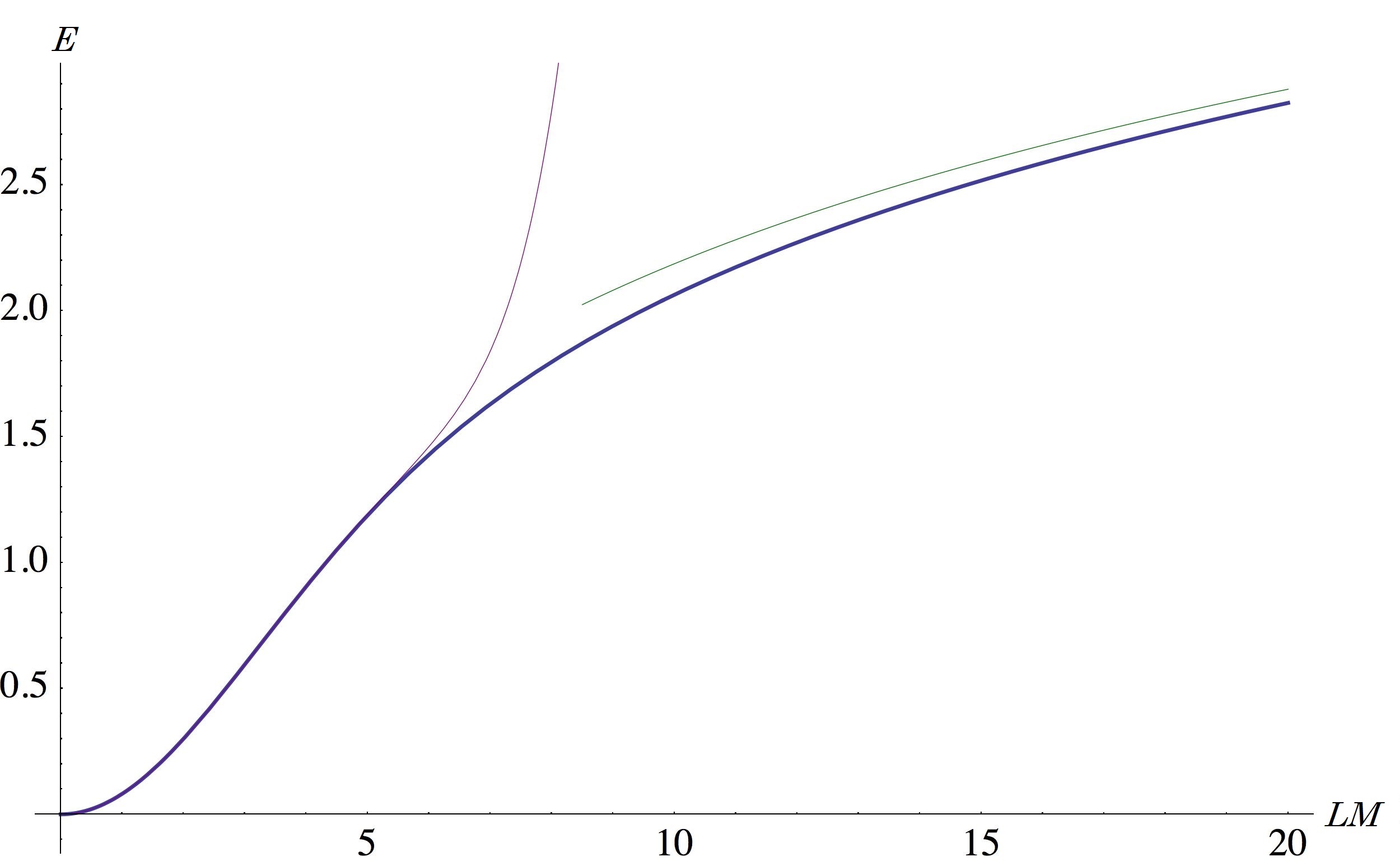}}
\caption{\label{fig:twist2E}\small The ground state ($S=0$) energy \eq{den1} as a function of the light-cone momentum $M$. The blue curve shows the numerical results. The purple curve is the 14th order small-$M$ approximation \eq{dpert1} and the green curve is the large-$M$ asymptotic result \eq{Large-ME1}.}
\end{center}
\end{figure}

The solution of the eigenvalue equation \eq{eqsm} is given by well-studied spheroidal wavefunctions \cite{Bateman:1955uq,SpheroidalFunctions,osipov2013prolate} which are also implemented in Mathematica.
The solution we need is the prolate angular spheroidal wavefunction of the first kind\footnote{Equation \eq{eqsm} is transformed to the canonical spheroidal differential equation by the redefinition $z\to z/\dl$.},
\beq
	G(z)=PS_{S,0}\(\frac{\dl M}{2},\frac{z}{\dl}\)
\eeq
and $t_0$ can be expressed through the spheroidal eigenvalue,
\beq
	t_0=-\frac{1}{2}-\lambda_{S,0}\(\frac{\dl M}{2}\)
\eeq
Thus, we have explicitly constructed the wavefunctions, and it remains to extract the energy. Knowing the eigenfunction one can do this by just setting $z$ to some value in the integral equation \eq{schrc}. For instance, with $z=0$ and assuming $S$ to be even (so the wavefunction is even) we find after  a set of simple transformations
\beq
\label{den1}
	\Delta^{(1)}_{S}=\frac{\lambda}{2\pi^2}\int\limits_0^1
	\frac{dx}{x}\(1-\cos\(\frac{\dl Mx}{2}\)
	\frac{PS_{S,0}\left(\frac{\dl M}{2}\,,x\right)}{PS_{S,0}\left(\frac{\dl M}{2}\,,0\right)}\)
	\ .
\eeq
It is now straightforward to compute the energy numerically or to get the perturbative expansion in $\dl M$. As an example, for the deformed ground state with $S=0$ we find
\beqa
\label{dpert1}
	\Delta^{(1)}_{J=2,S=0}&=&
	\frac{\lambda}{4\pi^2} 
	\(\frac{\dl^2 M^2}{6}-\frac{\dl^4 M^4}{216}+\frac{\dl^6 M^6}{9720}-\frac{83 \dl^8
   M^8}{48988800}+\frac{\dl^{10} M^{10}}{55112400}
	\right.
	\\ \nonumber
	&&
	\left.
	-\frac{409 \dl^{12}
   M^{12}}{6789847680000}-\frac{2881 \dl^{14} M^{14}}{2138802019200000}+\dots
	\right) \ .
\eeqa
For the same state, the expansion of the transfer matrix eigenvalue reads
\beq
\label{t2res}
	t_0=-\frac{1}{2}+\frac{ \dl^2 M^2}{6}+\frac{\dl^4 M^4}{1080}-\frac{\dl^6 M^6}{136080}-\frac{13 \dl^8 M^8}{244944000}+\dots \ \ .
\eeq
Another interesting regime to study is the limit of large $\dl M$ which we will analyze in detail in  section \ref{sec:largeMB}. Here we just quote the result,
\beq
\label{Large-ME1}
	 \Delta^{(1)}=\frac{\lambda }{2\pi ^2}\(\ln\(\dl M\)-\ln 2+\gamma+O\(\frac{1}{\dl M}\)\)
	\ ,
\eeq
where $\gamma$ is the Euler constant.
In Figure \ref{fig:twist2E} we show the analytic predictions at small and large $\dl M$, together with the numerical solution at finite $\dl M$.

\section{Long operators and the Landau-Lifschitz model \label{lll}}

The structure of the non-relativistic anomalous dimensions as a function of $M \dl$ appears to take a different form in supergravity, where the anomalous dimension only depends on the combination\footnote{While we have not yet computed the supergravity prediction for the anomalous dimension of the $tr Z^J$ operators for $J$ small, the form \eqref{schmetpoinc} of the supergravity background, together with (\ref{relmumuhat}) implies that they can only depend on this particular combination. } $\l M^2 \dl^2$, and field theory, where the one-loop anomalous dimensions have an infinite expansion in $M^2 \dl^2$. Of course, the former expression holds at strong t'Hooft coupling, whereas the latter holds at weak coupling, so there is no reason for the answers to match. Nevertheless, it would be interesting to compute and match the anomalous dimensions in an overlapping regime of validity of these two expansions. 

One such regime is the large spin regime $J \r \infty$ with $\tilde \l = \l/J^2$ held fixed and small (see e.g. \cite{Frolov:2002av}). In this limit, the energy of a classical string solution has a Taylor  expansion in $\tilde \l$, with $\a'$ corrections suppressed as $1/\sqrt{\l}$. On the SYM side, this limit corresponds to the thermodynamic limit of an infinite spin chain, whose low-energy dynamics is well described by semiclassical coherent states. In \cite{Kruczenski:2003gt} it has been shown that in the undeformed theory one can directly recover the string worldsheet $\s$-model from the spin chain in the small $\tilde \l$ expansion, up to one loop. 

In the following, we will briefly review  the derivation of the one-loop string worldsheet Hamiltonian from coherent states in the $\sl2$ sector of $\N=4$ SYM following \cite{Stefanski:2004cw,Bellucci:2004qr}. Then, we will show how the twisted spin chain described above precisely reproduces the one-loop anomalous dimensions at large $J$ obtained from the supergravity calculation of the string energy \eq{dbmn}. 

\subsection{Coherent states in the $\sl2$ spin chain}

In this section, we will briefly review the results of \cite{Stefanski:2004cw} on the 
coherent-state representation of the $\sl2$ spin chain and its relationship to rapidly spinning strings in an $AdS_3 \times S^1$ subspace of $AdS_5 \times S^5$. 

The $\sl2$ coherent states are defined as
\be
| \vec n \rangle = \sqrt{1-|\zeta |^2} \,\,{\rm e}\,^{\zeta S^-}  \left|0\right\rangle,
\qquad 
\left\langle \vec n\right|=\left\langle 0\right|\,{\rm e}\,^{\bar{\zeta} S^+} \sqrt{1-|\zeta |^2}   .
\ee
where $S^i$ corresponds to the $\mathfrak{sl}(2)$ spin generator  and  $|0\rangle$ corresponds to the highest weight state of weight $1/2$ (see Appendix \ref{sec:appsl2}):
\be\label{spin1/2rep}
S^+ | 0\rangle=0 \;, \;\;\;\;\;\; S^0 |0\rangle=- \frac{1}{2}\,  |0\rangle.
\ee
The basic property of the coherent states is that  
\be\label{hyperboloid}
\langle \vec n | S^i | \vec n \rangle = - \frac{1}{2} n^i \;, \;\;\;\;\;\;  \vec n \cdot \vec n \equiv  n_0^2 - n_+ n_- = 1.
\ee
 The three-component vector $\vec{n}$ parametrizes a two-dimensional hyperboloid (to be understood as a constant time slice in AdS$_3 \in$ AdS$_5$) and its components are explicitly given by 
\be\label{vector-n}
n_0 = \frac{1+|\zeta|^2}{1-|\zeta|^2}\;,\;\;\;\;\; n_+ = -\frac{2\zeta }{1-|\zeta|^2}\;,\;\;\;\;\; n_- = -\frac{2\bar \zeta }{1-|\zeta|^2}\;.
\ee
The coherent state for the entire spin chain is defined as $| \vec n \rangle = \prod_\ell |\vec n_\ell \rangle$. The expectation value of the full one-loop Hamiltonian \eq{Gham} is

\be
\langle \vec n | \Gamma | \vec n \rangle = \frac{\l}{8\pi^2} \sum_{\ell=1}^J  \langle \vec n_\ell | \langle \vec n_{\ell+1} | h_{\ell,\ell+1} | \vec n_\ell \rangle | \vec n_{\ell+1} \rangle 
\ee
The expectation value was explicitly evaluated in \cite{Stefanski:2004cw}, and yields

\be
\langle \vec n | \Gamma | \vec n \rangle = \frac{\l}{8\pi^2} \sum_{\ell=1}^J  \ln \frac{1+\vec{n}_\ell \cdot \vec{n}_{\ell+1} }{2} =\frac{\l}{8\pi^2} \sum_{\ell=1}^J  \ln \left[ 1-\frac{1}{4} (\vec n_{\ell+1}-\vec n_\ell)\cdot  (\vec n_{\ell+1}-\vec n_\ell) \right]
\ee
Next, one takes the  $J \r \infty$ limit with $\l/J^2$ fixed.  Introducing the continuous variable $\s \in (0, J)$ such that

\be
\vec n_\ell = \left. \vec n(\s)\right|_{\s=\ell} \;,\;\;\;\;\; \Rightarrow \;\;\;\; \vec n_{\ell+1} - \vec n_\ell =  \p_\s \vec n(\s) + \O\left( \p_\s^2 \vec n \right)
\ee
where each additional  $\s$ derivative is of order $1/J$ and thus can be dropped.  In this low energy limit, only the first term in the expansion of the logarithm survives, and we obtain 
\be\label{LLHam}
H_{1-loop} = - \frac{\l}{32 \pi^2} \int_0^J d\s \, \p_\s \vec n \cdot \p_\s \vec n
=\frac{\lambda }{32\pi ^2}
\int_{0}^{J}d\sigma \,\left[
 \partial _\sigma n_+\,\partial _\sigma n_-
 -\left(\partial _\sigma n_0\right)^2
 \right].
\ee
The Poisson brackets of the $n^i$ are dictated by the original commutators of the $S^i$ variables, with the usual replacement $\{\, ,\}_{P.B.} \r -i  [\,,]$. Taking into that $S^i\rightarrow -n^i/2$, we find from  (\ref{sl2-commrel}):
\begin{eqnarray}\label{Poisson}
 \left\{n_0(\sigma ),n_\pm(\sigma ')\right\}&=&\pm 2in_\pm(\sigma )\delta \left(\sigma -\sigma '\right)
\nonumber \\
\left\{n_+(\sigma ),n_-(\sigma ')\right\}&=&-4in_0(\sigma )\delta \left(\sigma -\sigma '\right).
\end{eqnarray}
The Hamiltonian equations for (\ref{LLHam}), (\ref{Poisson}) are the Landau-Lifshitz equations for the classical one-dimensional (non-compact) ferromagnet:
\begin{eqnarray}
 \partial _tn_0&=&\frac{i\lambda }{16\pi ^2}\left(n_-\partial _\sigma ^2n_+
 -n_+\partial _\sigma ^2n_-\right)
\nonumber \\
\partial _tn_\pm&=&\pm\frac{i\lambda }{8\pi ^2}\left(n_0\partial _\sigma ^2n_\pm-n_\pm\partial _\sigma ^2n_0\right)
\end{eqnarray}
It is easy to check that time evolution preserves the constraint (\ref{hyperboloid}) and thus only two of the three equations for the components of $\vec{n}$ are independent.

\subsection{Twisted boundary conditions}

The ground state of the spin chain corresponds to the solution $n_0=1$, $n_\pm=0$. It obviously has zero momentum. The total light-cone momentum according to (\ref{momentum-identif}) and the substitution $S_i\rightarrow -n_i/2$ is identified with
\begin{equation}
 P_-=\frac{i}{2}\int_{0}^{J}d\sigma \,n_-.
\end{equation}
Again it is easy to check using the Landau-Lifshitz equations that the momentum is conserved. The solution of the undeformed model that describes the ground state with momentum $M$ is given by
\begin{equation}\label{unde-gs}
 n_0=1,\qquad n_-=-\frac{2iM}{J}\,,
\end{equation}
which satisfies both the equations of motion and  the constraint in (\ref{hyperboloid}) because $n_+=0$. The solution has zero energy as expected. 

This solution is a complex saddle-point of the Landau-Lifshitz path integral. The operators $S^+$ and $S^-$ are conjugate in the representation (\ref{spin1/2rep}), and likewise the components $n_\pm$ of the vector $\vec{n}$   are complex conjugate by construction.  A field configuration with definite momentum but zero $S^+$ requires an analytic continuation in which $n_+$ and $n_-$  are treated as independent variables.

According to (\ref{bc-spinchain}), the dipole deformation results in the following twisted boundary conditions for the spin operators (recall that the total $R$-charge of the spin chain is equal to its length $J$ and $P_-$ is identified with $-iS^-$):
\begin{equation}
 S^i_{J+1}=\,{\rm e}\,^{-\dl JS^-_1}S_1^i\,{\rm e}\,^{\dl JS^-_1}.
\end{equation}
It is perhaps necessary to mention that the similarity transformation on the right-hand-side is not unitary, in the conventional norm of the spin 1/2 $\mathfrak{sl}(2)$ representation. The momentum operator in this norm is not Hermitian.

The explicit form of the similarity transformation can be most easily worked out in the differential-operator representation (\ref{spin-s}):
\begin{eqnarray}
 S^-_{J+1}&=&S_1^-
\nonumber \\
S^0_{J+1}&=&S_1^0-\dl JS_1^-
\nonumber \\
S^+_{J+1}&=&S_1^+-2\dl JS_1^0+\dl^2J^2S_1^-.
\end{eqnarray}
Accordingly, the spin variable in the coherent-state path integral will obey in the quasi-periodic boundary conditions:
\begin{eqnarray}
 n_-(t,\sigma +J)&=&n_-(t,\sigma )
\nonumber \\
n_0(t,\sigma +J)&=&n_0(t,\sigma )-\dl Jn_-(t,\sigma )
\nonumber \\
n_+(t,\sigma +J)&=&n_+(t,\sigma )-2\dl Jn_0(t,\sigma )+\dl^2J^2n_-(t,\sigma ).
\end{eqnarray}
Field configurations satisfying these boundary conditions are necessarily complex.
The ground state solution, for example, is now $n_0=1$, $n_+=-2\dl \sigma $. The solution still has zero energy. The ground state at non-zero momentum gets modified in a more complicated way. 

To find such a solution, we try the Ansatz 
\begin{eqnarray}\label{ansatz}
 n_-&=&-\frac{2iM}{J}
\nonumber \\
n_0&=&\alpha +\beta \sigma ,
\end{eqnarray}
where $\alpha $ and $\beta $ may depend on time. We need not make assumptions about  the plus component of $\vec{n}$, it  is simply determined by the constraint (\ref{hyperboloid}). 

The $n_-$ equation is identically satisfied by the Ansatz (\ref{ansatz}), while the $n_0$ equation becomes
\begin{equation}
 \partial _tn_0=\frac{i\lambda }{16\pi ^2}\,\partial _\sigma ^2n_0^2,
\end{equation}
where we used the constancy of $n_-$ and the constraint (\ref{hyperboloid}) to eliminate $n_+$. From this equation we find that $\beta $ is time-independent and 
\begin{equation}
 \alpha =\frac{i\lambda }{8\pi ^2}\,\beta ^2t.
\end{equation}
The boundary conditions (again we only need to check two equations out of three) then fix $\beta $:
\begin{equation}
 \beta =\frac{2i\dl M}{J}\,.
\end{equation}
We can now compute the energy (equal to the one-loop anomalous dimension) from (\ref{LLHam}):
\begin{equation}
 \Delta^{(1)} =-\frac{\lambda }{32\pi ^2}\,J\beta ^2=\frac{\lambda \dl^2M^2}{8\pi ^2J}\,,
\end{equation}
which agrees exactly with the string prediction (\ref{dbmn}) expanded to the first order in $\lambda $, when the R-charge $R$ is identified with $J$.

\section{One-loop spectrum from the Baxter equation \label{baxter}}

In the previous sections we have shown how to obtain the 1-loop spectrum in the $\sl2$ sector for $J=2$ and in the large $J$ semiclassical limit. In this section we will use the full power of integrability to access the spectrum at  any $J$ by solving the twisted spin chain described in section \ref{sec:3last}.

\subsection{The Baxter equation}

As discussed in section \ref{sec:3last}, in our model it is impossible from the very beginning to obtain the spectrum from the conventional coordinate or algebraic Bethe Ansatz. The reason for this is the absence of a suitable pseudovacuum reference state. Instead, we will use the Baxter equation approach, closely linked with Sklyanin's separation of variables \cite{Sklyanin:1991ss, Sklyanin:1995bm}.

For generic $J$ the spin chain Hamiltonian is a rather involved integral operator in $J$ variables, making it nearly impossible to directly compute its eigenvalues. The separation of variables approach reduces the problem to a set of decoupled difference equations. In our case we expect them to coincide with the Baxter equation, which for models based on the rational $\sl2$ R-matrix has the general form \cite{Sklyanin:1991ss}
\beq
\label{BaxD}
	\Delta_+(u)Q(u+i)+\Delta_-(u)Q(u-i)=t(u)Q(u)
\eeq
Here $\Delta_{\pm}(u)$ are some model-dependent functions, while $t(u)$ is the eigenvalue of the trace of the monodromy matrix. In particular, for the undeformed
${sl}(2)$ spin chain the Baxter equation has been derived in \cite{Derkachov:2002tf} by implementing the full separation of variables program. Below we will argue that for our deformed chain the equation remains the same.

For the undeformed theory, which corresponds to an $\sl2$ spin chain with $J$ sites, the Baxter equation reads
\beq
\label{Baxund}
	(u+i/2)^JQ(u+i)+(u-i/2)^JQ(u-i)=t(u)Q(u)
\eeq
Both the Q-function $Q(u)$ and the transfer matrix eigenvalue $t(u)$ are fixed by the Baxter equation and the requirement that they be  polynomials in $u$.
The 1-loop energy can then be extracted from the Q-function via 
\beq
\label{EQ}
	\Delta^{(1)}=\frac{i\lambda}{8\pi^2}\left.\partial_u\log\frac{Q(u+i/2)}{Q(u-i/2)}\right|_{u=0}
\eeq
This expression is equivalent to the usual Bethe ansatz formula given by the sum over Bethe roots $u_j$ which are the zeros of the Q-function
\beq
	\Delta^{(1)}=\frac{\lambda}{8\pi^2}\sum_{j=1}^S\frac{1}{u_j^2+1/4}
\eeq
with
\beq
	Q(u)=\prod_{j=1}^S(u-u_j)\ \ .
\eeq

Even though our spin chain model has twisted boundary conditions, we propose that the Baxter equation has the same form \eq{Baxund}. While we have not proven this statement,
let us describe an algebraic argument strongly supporting this proposal\footnote{F. L.-M. is grateful to Stefano Negro for related discussions.}. In any integrable model described by the Baxter equation \eq{BaxD}, one can bring it to the form
\beq
\label{Bax01}
	Q_0(u+i)-t(u)Q_0(u)+ W(u-i/2)Q_0(u-i)=0
\eeq
where
\beq
\label{ddpm}
	W(u)=\Delta_+(u-i/2)\Delta_-(u+i/2)\ 
\eeq
and $Q_0(u)$ is related to the original Q-function by
\beq
	Q_0(u)=Q(u)/f(u)
\eeq
with $f(u)$ being a solution of the simple equation
\beq
	f(u+i)=f(u)/\Delta_+(u)\ .
\eeq
The advantage of this redefinition is that the quantity $W(u)$ appearing in \eq{Bax01} has a clear algebraic interpretation -- as discussed in  \cite{Sklyanin:1991ss} it should be the quantum determinant of the monodromy matrix $T(u)$. In general for a monodromy matrix having the form
\beq
	T(u)=\begin{pmatrix} T_{11}(u) & T_{12}(u) \\ T_{21}(u) & T_{22}(u) \end{pmatrix} \ ,
\eeq
the quantum determinant is defined as
\beq
\label{qddef}
	W(u)=T_{11}(u+i/2)T_{22}(u-i/2)-T_{12}(u+i/2)T_{21}(u-i/2) \ .
\eeq
While it is an operator acting on the Hilbert space of the model, it commutes with the entries $T_{ij}$ and acts as a scalar in irreducible representations, so in \eq{Bax01} we understand $W(u)$ as a scalar function. We see that both coefficients $t(u)$ and $W(u)$ in \eq{Bax01} are well defined algebraic objects. Therefore one expects that in this form the equation is valid for all models\footnote{We also assume that the model can be described at all by a homogenous 2nd order Baxter equation. This is not the case e.g. in some \textit{open} deformed spin chains where the Baxter equation is not homogenous \cite{Cao:2013nza}.}
that are based on the rational $sl(2)$ R-matrix \footnote{Let us also mention that since the equation \eq{Bax01} includes the trace and the (quantum) determinant of the $2\times 2$ matrix $T(u)$, it may be seen as a kind of eigenvalue equation, and interpreted as a quantum version of the classical spectral curve. See e.g. \cite{Chervov:2007bb,Chervov:2006xk} for a detailed discussion of this point of view.}.

As a check, for the undeformed $sl(2)$ spin chain the usual Baxter equation \eq{Baxund} with
\beq
\label{dpm0}
	\Delta_+(u)=(u+i/2)^J, \ \ \  \ \Delta_-(u)=(u-i/2)^J
\eeq
can be brought to the universal form \eq{Bax01} with $W(u)$ read off from \eq{ddpm}, 
\beq
	W(u)=(u+i/2)^J (u-i/2)^J\ .
\eeq
As expected this expression is precisely the quantum determinant of the monodromy matrix \eq{T0sl2}, which can be verified by a direct calculation. 

In order to find the Baxter equation for our deformed spin chain it remains only to compute the quantum determinant of the twisted monodromy matrix \eq{Ttw} entering the Baxter equation \eq{Bax01}. Plugging in its entries given in \eq{Ttw} into the definition \eq{qddef} we see that all the dependence on the deformation parameter $\dl$ cancels! This means that the quantum determinant is unchanged so the Baxter equation should be the same as before, and
may be written as either \eq{Bax01} or in the conventional form \eq{Baxund}.

\subsection{Asymptotics and analyticity of the Q-functions}

While the Baxter equation remains the same as in the undeformed case,
\beq
	(u+i/2)^JQ(u+i)+(u-i/2)^JQ(u-i)=t(u)Q(u)\ ,
\eeq
a crucial new feature is that the Q-functions are no longer polynomials of $u$. This follows from the fact that the transfer matrix eigenvalue $t(u)$ (which remains a polynomial) now includes a $u^{J-1}$ term with a nonzero coefficient fixed by the deformation parameter
\beq\label{tugen}
	t(u)=2u^J+\dl MJu^{J-1}+\dots
\eeq
This can be seen directly from the expression for the transfer matrix
\beq
	{\rm tr} \;T(u)=A(u)+D(u)+\dl J B(u)
\eeq
taking into account that the leading coefficient of $B(u)$ is proportional to $S_-$ and thus acts diagonally on the states with a definite value of $M$. From the Baxter equation \eq{Baxund} we now see that the Q-functions cannot even have powerlike asymptotics at large $u$, much less be polynomials. Instead we deduce from \eq{Baxund} that e.g. for $J=2$, the asymptotic behaviour of the two solutions $Q_1$ and $Q_2$ reads\footnote{Let us recall that in the undeformed case the two Q-functions have asymptotics $u^S$ and $u^{-S-1}$ for integer $S$.}
\beq
\label{Qlarge}
	Q_j(u)\sim  u^{-3/4}e^{\alpha_j\sqrt{u}}\(1+\sum\limits_{n=1}^\infty\frac{c_{n}}{(\alpha_j)^nu^{n/2}}\),\ \ \ u\to\infty
\eeq
where
\beq
	\alpha_{1,2}=\pm i\sqrt{8\dl M}
\eeq
and the coefficients $c_n$ do not depend on $j$. The asymptotics has this peculiar form both for the ground state and the excited states. For higher $J$ the asymptotic behaviour is also similar.

Our results for the Q-function asymptotics should provide guidance for generalizing to this setting the Quantum Spectral Curve that captures the all-loop spectrum of local operators in the undeformed $\cN=4$ SYM \cite{Gromov:2013pga,Gromov:2014caa}. 
Curiously, asymptotic behaviour of the same type was found in \cite{Gromov:2016rrp} for the Q-functions of the Quantum Spectral Curve describing the quark-antiquark potential in undeformed $\cN=4$ SYM.

In the undeformed case, requiring that $Q(u)$ and $t(u)$ be polynomials fixes both of them and allows to extract the 1-loop energy from the relation \eq{EQ}
\beq
\label{DQ}
	\Delta^{(1)}=\frac{i\lambda}{8\pi^2}\left.\partial_u\log\frac{Q(u+i/2)}{Q(u-i/2)}\right|_{u=0}
\eeq
In the deformed case, the Q-function clearly cannot be polynomial. However we observed that requiring it to be free of singularities at finite $u$ completely fixes both the Q-function and the transfer matrix eigenvalue! Moreover, the energy is then given by the same expression \eq{DQ}. We have not proven these properties, but have checked extensively that this procedure works in explicit calculations to high order at small $\dl$, described below. In general the regularity of the Q-function and the expression for the energy should follow from a complete implementation of the SoV program as was done in \cite{Derkachov:2002tf} for the undeformed case. We leave this question for future work\footnote{Let us note that some of the key ingredients in the SoV program remain unchanged in the deformed model. For example, the separated coordinates $x_k$ are defined as the operator zeros of the element $T_{12}(u)$ of the monodromy matrix, and this element is not modified under the deformation. Moreover the conjugated variables to these coordinates are given by $T_{11}(x_k)$ and $T_{22}(x_k)$ and are also the same, as follows from \eq{Ttw} since $T_{12}(x_k)=0$.}.

Let us mention that $\dl$ appears in the Baxter equation approach solely in \eq{tugen}, where it is multiplied by $M$. Thus, as expected, the energy is a function only of the combination $\dl M$.

In summary, we have presented the Baxter equation which should capture the spectrum of all operators in the $\sl2$ sector. In the next section we will use it to compute the anomalous dimensions as a series in the deformation parameter.

\subsection{Perturbative solution}
\label{sec:pertq}

\textbf{Algorithm.\ \ } Let us describe how one can solve the Baxter equation and compute the energy as an expansion in $\dl M$. Our strategy is to build two Q-functions perturbatively in $\dl M$ and require the existence of a solution without singularities. While we only need the regular Q-function, it is instructive to construct both of them. We start from the two leading order Q-functions and then build the solutions at higher orders by using the iterative algorithm developed for the $\cN=4$ SYM Quantum Spectral Curve and described in
\cite{Gromov:2015vua}. It is based on the method of variation of constants. As expected we find that the solution is given in terms of $\eta$-functions like in \cite{Marboe:2014gma, Gromov:2016rrp} defined as
\beq
	\eta_{s_1,\dots,s_k}(u)=\sum\limits_{n_1>n_2>\dots>n_k\geq0}^\infty\frac{1}{(u+in_1)^{s_1}\dots (u+in_k)^{s_k}}
\eeq
We used Mathematica packages for working with these functions given in \cite{Gromov:2015vua, Gromov:2015dfa} (see also \cite{Marboe:2014gma}) which allow to construct the high-order iterative solution with ease.


For example, when $J=2$ the two Q-functions for the ground state at $\dl=0$ are
\beq
	q_1(u)=1, \ \ q_2(u)=\eta_2^+
\eeq
where we denote
\beq
	f^{\pm}=f(u\pm i/2),\ \ \ f^{[+a]}=f(u+ia/2)
\eeq
Writing the transfer matrix eigenvalue $t(u)$ as a series in $\dl M$
\beq
\label{t0exp}
	t(u)=2u^2+2\dl Mu+t_0 \ , \ \ \ 
	t_0=-\frac{1}{2}+\tau_1 \dl M  + \tau_2 (\dl M)^2+\dots
\eeq
and solving the Baxter equation iteratively we find two solutions given by
\beqa
	q_1(u)&=&1+\dl M\(i\tau_1{\eta_1^+} -u-i/2\) \\ \nonumber
	&+&\frac{\dl^2 M^2}{6}\[\vphantom{\frac{1}{2}} -6\tau_1^2(\eta_2^++\eta_{1,1}^+)
	-i(-6\tau_2+1+3\tau_1(-2\tau_1+2u+i))\eta_1^+
	\right.
	\\ \nonumber
	&&
	\left.
	\vphantom{\frac{1}{2}}
	-6\tau_1(u-i)+2u^2+3iu-1
	\] +\ O(\dl^3 M^3) \ \ \ , \\ \nonumber
	q_2(u)&=&\eta_2^++O(\dl M)
\eeqa
The true solution is a linear combination of these two functions and it should be free of singularities. It is clear that $q_2(u)$ cannot enter this linear combination (it has poles already at the leading order which cannot cancel those in $q_1(u)$). Then, requiring that $q_1(u)$ is regular fixes immediately the coefficients 
\beq
	\tau_1=0,\ \ \tau_2=1/6
\eeq
in complete agreement with the perturbative result \eq{t2res} we presented in section 4. The two solutions then read
\beqa
\label{q1p}
	q_1(u)&=&1+\dl M \left(-u-\frac{i}{2}\right)+\frac{1}{6} \dl^2 M^2 \left(2 u^2+3 i
   u-1\right) +O(\dl^3 M^3)\\ \nonumber
	q_2(u)&=&\eta_2^+
	+\dl M\(2\eta_1^+ + \frac{1}{2}(-2u+i)\eta_2^+\)\\ \nonumber
	&&+ \dl^2 M^2\((-2u+i)\eta_1^++\frac{1}{6}(u-i)(2u-i)\eta_2^++\frac{7}{6}(1+2iu)\)
	\\ \nonumber &&
	+O(\dl^3 M^3)
\eeqa
Extracting the 1-loop energy from $q_1(u)$ via \eq{DQ} we get
\beq
	\Delta^{(1)}=\frac{\lambda}{4\pi^2} \(\frac{\dl^2 M^2}{6}+O(\dl^3 M^3)\)
\eeq
in agreement with our direct perturbative calculation \eq{dpert1}. 

Proceeding to higher orders we fix one by one the coefficients $\tau_n$ appearing in \eq{t0exp} and also fix the Q-function itself. We have checked to $\dl^8$ order that the transfer matrix eigenvalue obtained in this way matches \eq{t2res} and the energy also matches the result obtained by perturbatively diagonalizing the Hamiltonian \eq{schrc}.

We have also checked the matching of the energy with direct diagonalization of the Hamiltonian to first several orders for the first two excited states with $J=2$. This gives strong support to our guess that the Baxter equation is the same, the true Q-function should be regular and the energy is given by the same formula \eq{DQ}. 

Let us note that at each order in $\dl M$, the Q-function \eq{q1p} is simply a polynomial. The peculiar asymptotic behaviour at large $u$   \eq{Qlarge} is irrelevant in the regime $u \ll 1/(\dl M)$ that we study here, and becomes visible only when $u\gg 1/(\dl M)$. It is still possible to link our perturbative solution with the large $u$ asymptotics by considering separately the intermediate regime $u \sim 1/(\dl M)$, similarly to what was done in the perturbative calculation of \cite{Gromov:2016rrp} for the quark-antiquark potential. Luckily we do not need to go through that rather nontrivial procedure, as in our case the Q-function is completely fixed by the regularity condition. Nevertheless, for completeness we present the solution at the intermediate scale in Appendix \ref{sec:scales}.

Noting that the Q-function is a polynomial whose degree is growing with the order of perturbation theory in $\dl M$, we can bypass going through the iterative procedure described above, which provides also the other solution that we don't actually need. We can simply make an Ansatz for the Q-function as a polynomial and find that all of its coefficients are fixed by the Baxter equation together with $t(u)$, order by order in $\dl M$. 
\bigskip

\par \noindent
\textbf{Results.\ \ } Using the Baxter equation we can now easily study states with $J>2$.
It is especially interesting to explore the large $J$ limit in order to compare the energy with the string prediction \eq{dbmn}. Computing the ground state energy for $J=2,3,\dots,14$ we found that the first several coefficients are given by 
\beq
\label{D1J}
	\Delta^{(1)}=\frac{\lambda}{4\pi^2}\(
	\frac{1}{2(J+1)}\dl^2M^2-\frac{1}{24 (J+1)^2}\dl^4 M^4
	+\frac{J^2+J+2}{720 (J+1)^3 (J+2)}\dl^6 M^6
	+O(\dl^8 M^8)\)
\eeq
At the same time, for large $J$ the string prediction \eq{dbmn} (in which the R-charge $R$ is identified with $J$) gives 
\beq
	\Delta=J+\frac{\lambda}{8\pi^2}\frac{\dl^2 M^2}{J}+\dots
\eeq
in perfect agreement with the large $J$ expansion of the spin chain result \eq{D1J}~!
This provides a nontrivial test of the holographic duality for the Schr\"{o}dinger background, supplementing the calculation from the Landau-Lifshitz model described above. 

In order to match the classical string prediction, only the $\dl^2 M^2$ term in the spin chain result is important. We expect that terms with higher power of $\dl M$ in \eq{D1J} (as well as higher-loop corrections at weak coupling) should similarly play a role in reproducing quantum corrections to the string prediction.

We have also computed to high order the expansion of the 1-loop energy for the ground state at $J=2,3$ and $4$. For $J=2$ we reproduced the result \eq{dpert1} obtained in section 4, while for $J=3,4$ we get
\beqa
	\Delta^{(1)}_{J=3}&=& 
	\frac{\lambda}{4\pi^2} \(
	\frac{\dl^2 M^2}{8}-\frac{\dl^4 M^4}{384}+\frac{7 \dl^6 M^6}{115200}-\frac{13 \dl^8
   M^8}{10321920}
	\right. \\ \nonumber  && \left.
	+\frac{3097 \dl^{10} M^{10}}{162570240000}+O\left(\dl^{12}M^{12}\right)\)
	\\ \nonumber
	\\ \nonumber
	\Delta^{(1)}_{J=4}&=&
	\frac{\lambda}{4\pi^2} 
	\(
	\frac{\dl^2 M^2}{10}-\frac{\dl^4 M^4}{600}+\frac{11 \dl^6 M^6}{270000}-\frac{533 \dl^8
   M^8}{529200000}
	\right. \\ \nonumber  && \left.
	+\frac{4813 \dl^{10} M^{10}}{238140000000}+O\left(\dl^{12}M^{12}\right)\)
\eeqa
It is certainly straightforward to get even more terms. 
Finally, we have obtained the leading term in the 1-loop anomalous dimension
 for any excited state with $J=2$. Identifying the states by their spin $S$ in the undeformed theory, we have\footnote{We remind the reader that $h(S)$ denotes the $S$-th harmonic number.}
\beq
	\Delta^{(1)}_{J=2,S}=\frac{\lambda}{4\pi^2}\(2h(S)-\frac{\dl^2 M^2}{2 (2 S-1) (2 S+3)}+O(\dl^4M^4)\)
\eeq
To get this result we computed the energy for $S=0,1,\dots,12$  and observed that the leading coefficient is captured by the simple expression above. Notice that the $\dl^2 M^2$ correction does not contain the BFKL pole at $S=-1$ present in the leading order term $2h(S)$.

\subsection{Exact solution for $J=2$ and large $\dl M$ expansion}
\label{sec:largeMB}

In this section we will show that for $J=2$, the Baxter equation can be recast as a differential equation, allowing in particular to find the behavior of the energy at large $\dl M$.

\bigskip

\par \noindent
\textbf{Differential equation}. \ \ \ The Baxter equation for $J=2$ can be reduced to a 2nd order differential equation via Mellin transform  \cite{Faddeev:1994zg}. Namely, we represent the Q-function as
\begin{equation}\label{Mellin}
 Q(u)=\cosh \pi u\int_{0}^{1}dz\,z^{-iu-\frac{1}{2}}\left(1-z\right)^{iu-\frac{1}{2}}\hat{Q}(z).
\end{equation}
Assuming that $\hat{Q}(z)$ is analytic on the interval $(0,1)$ and has no singularities at the endpoints, this formula defines a function analytic in the finite part of the complex plane. The integral converges on the strip $-1/2<\mathop{\mathrm{Im}}u<1/2$. Analytic continuation from the strip into the whole complex plane leads to simple poles at $u=in-i/2$, but those are canceled by the cosh prefactor. It is important to emphasize that $\hat{Q}(z)$ should not have any singularities at $z=1$ and $z=0$ for these analyticity properties to hold.

Substitution of (\ref{Mellin}) into the Baxter equation \eq{Baxund} with $t(u)$ given by the second-order polynomial (\ref{t0exp}) converts it into the second-order differential equation\footnote{For the $J$-site spin chain the resulting differential equation would contain up to order-$J$ derivatives.}:
\begin{equation}\label{Heun-equation}
 z(1-z)\hat{Q}''
 +\left[2i\dl Mz(1-z)
 +1-2z\right]\hat{Q}'
 +i\dl M\left(1-2z\right)\hat{Q}=
 (t_0+1/2)\hat{Q}.
\end{equation}
A further change of variables 
\begin{equation}
 \hat{Q}(z)=\,{\rm e}\,^{-i\dl Mz}F (2z-1)
\end{equation}
reduces (\ref{Heun-equation}) to the self-conjugate form. The result is the spheroidal differential equation:
\begin{equation}\label{spheroidalWeq}
 \left[
 \frac{d}{dx}\left(1-x^2\right)\frac{d}{dx}+\frac{1}{4}\,\dl ^2M^2\left(1-x^2\right)-(t_0+1/2)
 \right]F (x)=0.
\end{equation}
This is precisely the equation which we obtained from the eigenvalue equation for the transfer matrix in section 4, see \eq{eqsm} (replacing in the latter equation $z\to z/\dl$ yields \eq{spheroidalWeq}).

The differential equation (\ref{spheroidalWeq}) defines an eigenvalue problem. The variable $x$ takes values on the interval $(-1,1)$, and we need to impose the boundary conditions on $F(x)$ at the endpoints. The points $x=-1$ and $x=1$ are regular singular points, and the general solution of the equation behaves as
\begin{eqnarray} 
 F(x)&\simeq& \mathbf{C}_1+\mathbf{C}_2\ln (1+x)\qquad (x\rightarrow -1)
\nonumber \\
F(x)&\simeq& \mathbf{K}_1+\mathbf{K}_2\ln(1-x)\qquad (x\rightarrow 1),
\end{eqnarray}
potentially having logarithmic branch points there. The Mellin transform would convert the log cuts into singularities of $Q(u)$ at half-integer points on the imaginary axis, and therefore we need to demand that
\begin{equation}
 \mathbf{C}_2=0,\qquad \mathbf{K}_2=0,
\end{equation}
which forms an overcomplete set of boundary conditions. They define an eigenvalue problem for $t_0$.
The solution is the prolate angular spheroidal wave function of the first kind \cite{Bateman:1955uq,SpheroidalFunctions,osipov2013prolate}, just as we found in section \ref{sl2sec},
\begin{equation}
 F (x)=PS_{S,0}\left(\frac{\dl M}{2}\,,x\right)
\end{equation}
and $t_0$ is written in terms of the spheroidal eigenvalue:
\begin{equation}
 t_0=-1/2-\lambda _{S,0}\left(\frac{\dl M}{2}\right).
\end{equation}
Finally, from the formula for the energy \eq{DQ} in terms of the Baxter function (taking $S$ to be even for simplicity, so $F(x)$ is also even) we get
\begin{eqnarray}\label{Twist2energy(S)}
\Delta^{(1)}&=&\frac{\lambda }{2\pi ^2}\int_{0}^{1}\frac{dx}{1-x^2}\,
 \left[1-
 \left(\cos\frac{\dl M}{2}\,\cos\frac{\dl Mx}{2}+x\sin\frac{\dl M}{2}\,\sin\frac{\dl Mx}{2}\right)
\right. \nonumber \\
&&\left.\times 
\frac{PS_{S,0}\left(\frac{\dl M}{2}\,,x\right)}{PS_{S,0}\left(\frac{\dl M}{2}\,,1\right)}
 \right].
\end{eqnarray}
This formula should be equivalent to the one following directly from the Hamiltonian we presented in \eq{den1} in section \ref{sl2sec}.

\bigskip

\par \noindent
\textbf{Large $\dl M$ asymptotics.} \ \ \ The large $\dl M$ regime corresponds to the semiclassical approximation in the differential equation (\ref{spheroidalWeq}). The leading semiclassical approximation for the ground-state energy is
\begin{equation}
t_0\simeq \frac{\dl^2M^2}{4}\,,
\end{equation}
while the semiclassical wave function is given by
\begin{equation}
 F (x)\simeq \,{\rm e}\,^{\frac{\dl M}{2}\,\sqrt{1-x^2}}.
\end{equation}
The semiclassical approximation breaks down near the turning points $x=\mp 1$.
In their vicinity, for $1\mp x\sim 1/\dl^2M^2$, the equation can instead be linearized in $1\pm x$:
\begin{equation}
 \left[2\,\frac{d}{dx}\left(1\pm x\right)\frac{d}{dx}-\frac{\dl^2M^2}{4}\right]F =0,
\end{equation}
and solved in terms of modified Bessel function:
\begin{equation}
F (x)\simeq I_0\left(\dl M\sqrt{\frac{1\pm x}{2}}\right).
\end{equation}
The Mellin transform of the Baxter function in the semiclassical approximation therefore is
\begin{equation}
 \hat{Q}(z)\simeq \,{\rm e}\,^{\dl M\left[-iz+\sqrt{z(1-z)}\right]},
\end{equation}
 except for when  $z$ is very small, $z\sim 1/\dl^2M^2$. In this case,
\begin{equation}\label{Qapprox}
 \hat{Q}(z)\simeq I_0\left(\dl M\sqrt{z}\right).
\end{equation}
The two expressions match in the overlapping region of validity $1/\dl M\gg z\gg 1/\dl^2 M^2$.

The integrand in (\ref{Mellin}) at large $\dl M$ rapidly oscillates, and one might think that the  saddle-point approximation for the integral will result in an exponentially small Baxter function. This is not the case, because the contour of integration passes through the turning points where the semiclassical approximation breaks down. In order to apply the saddle-point method, the contour of integration has to be extended from $-\infty $ to $+\infty $. This can be achieved by pulling out a backtracking loop from $0$ to $-\infty $, and a similar one from $1$ to $+\infty $. The oscillating integral from $-\infty $ to $+\infty $ is exponentially small and can be dropped, while the integral along the segments $(0,-\infty )$ and $(1,+\infty )$ and is dominated by the endpoint region of $z$'s close to zero and to one:
\begin{equation}
 Q(u)\simeq \cosh \pi u \int_{0}^{-\infty }dz\,z^{-iu-\frac{1}{2}}\hat{Q}(z)+\left(u\rightarrow -u\right).
\end{equation}
Using the approximate solution (\ref{Qapprox}) in the near-endpoint region we find:
\begin{equation}
 Q(u)\simeq \cosh \pi u \int_{0}^{\infty }dz\, z ^{-iu-\frac{1}{2}}J_0\left(\dl M\sqrt{z}\right)+\left(u\rightarrow -u\right),
\end{equation}
which evaluates to
\begin{equation}
 Q(u)\simeq \cosh\pi u
 \left(\frac{\dl M}{2}\right)^{2iu-1}\frac{\Gamma \left(-iu+\frac{1}{2}\right)}{\Gamma \left(iu+\frac{1}{2}\right)}
 +\left(u\rightarrow -u\right).
\end{equation}
Substituting this into \eq{DQ} we get for the energy at large $\dl M$:
\begin{equation}\label{Large-ME}
 \Delta^{(1)}=\frac{\lambda }{2\pi ^2}\,\ln\frac{\dl M\,{\rm e}\,^{\gamma }}{2}\,,
\end{equation}
where $\gamma $ is the Euler constant. This is the result we presented without derivation in  equation \eq{Large-ME1} in section \ref{sl2sec}.

\section{Discussion and future directions \label{concl}}

In this paper we have demonstrated the power of integrability in an application to a deformed version of $\cN=4$ SYM, where the deformation mixes the internal and the spacetime symmetries of the theory. 
For the first time, we have obtained the 1-loop spectral equations describing the nonprotected operator spectrum for a large subsector in a model of this type (namely, for the $\sl2$ subsector). The deformation profoundly affects the integrable structure, rendering the usual Bethe ansatz inapplicable from the very start. We proposed that the spectrum is instead captured by a Baxter equation whose solutions are non-polynomial, yet regular. We obtained the $J=2$ spectrum in terms of known special functions. We also presented explicit results at higher $J$ as a series in the deformation parameter. In the large $J$ limit, we showed that they reproduce the classical string theory prediction that we computed independently, providing a quantitative test of the holographic duality for the Schr\"odinger background. We have confirmed the matching by solving the corresponding Landau-Lifschitz effective theory.

The calculations we presented serve as a first step towards the derivation of the full quantum spectral curve (QSC) \cite{Gromov:2013pga,Gromov:2014caa} describing the all-loop spectrum for generic operators. Based on experience with other deformations of $\cN=4$ SYM \cite{Gromov:2015dfa,Kazakov:2015efa,Gromov:2016rrp,Gromov:2017cja}, it is natural to expect that functional equations of the QSC will remain almost unchanged, with the deformation affecting primarily the asymptotic behaviour of the Q-functions. The highly surprising asymptotic behaviour we observed at one loop should provide key guidance for the complete set of asymptotics in the QSC. It remains an interesting problem to classify all possible deformations of the rather rigid QSC structure, and in particular to understand if the asymptotics we found (similar to those in \cite{Gromov:2016rrp}) can be incorporated in the geometrical construction of \cite{Kazakov:2015efa}.

Our integrability-based  results for the spectrum  rely on the conjectured solution of an $\sl2$ spin chain with a Jordan cell twist. It would be interesting to rigorously derive our Baxter equation by implementing the full separation of variables program for this model, following \cite{Derkachov:2002tf}. In absence of the usual Bethe Ansatz, this alternative method  would also provide a key handle on the the wavefunctions. This in turn would allow us to  explore the three-point correlators in this model, which have previously been also studied holographically \cite{Volovich:2009yh, Fuertes:2009ex} and were shown to take a very particular form.

A key feature of string theory in  Schr\"odinger backgrounds is the appearance of non-protected supergravity states whose anomalous dimensions are non-vanishing, but can remain small even at large 't Hooft coupling. We succeeded in reproducing the spectrum of such states in the BMN limit, both from the coherent-state formulation of the spin chain and from the approximate solution of the Baxter equation. These results are encouraging but they rely on the one-loop analysis on the field-theory side. All the experience with the undeformed $\mathcal{N}=4$ SYM teaches us that higher-order corrections, especially those beyond the wrapping order, are important to get a complete match with string-theory predictions at strong coupling.

As explained in the introduction, one important motivation for studying 
Schr\"{o}dinger holography is its relevance for the holographic description of extreme Kerr black holes. The Schr\"odinger backgrounds appearing in that context are deformations of $AdS_3$ rather than $AdS_5$. Exact string quantization on the  $AdS_3$ backgrounds supported by the NS-NS flux provides important clues \cite{Azeyanagi:2012zd}. Recent developments in the $AdS_3/CFT_2$ integrability \cite{Babichenko:2009dk,Sfondrini:2014via,Baggio:2017kza,Borsato:2016xns,Borsato:2015mma} may give us further insight into this very interesting subject, since the dipole deformation can be incorporated into the $AdS_3/CFT_2$ framework via the Drinfeld-Reshetikhin twist of its integrable structure, much like in the $AdS_5/CFT_4$ case  studied in this work.

\bigskip

\par \noindent
\textbf{Acknowledgements}
\smallskip

\noindent We would like to thank R.~Borsato, J.~Caetano, D.~Cassani, A.~Cavaglia, D.~Fioravanti, V.~Kazakov, G.~Korchemsky, I.~Kostov, L.~Martucci, J.~Minahan, S.~Negro, O.~Ohlsson~Sax and R.~Tateo for very interesting discussions.
F.~L.-M. and K.~Z. thank the Galileo Galilei Institute for Theoretical Physics (GGI) for hospitality during the program ``New Developments in AdS$_3$/CFT$_2$ Holography''. In addition, F.~L.-M. thanks INFN as well as the ACRI (Associazione di Fondazioni e di Casse di Risparmio S.p.a.) for partial support. F.~L.-M. is also grateful for hospitality to University of Torino where a part of this work was done.
K.~Z. would like to thank FEFU, Vladivostok and Centro de Ciencias de Benasque Pedro Pascual for hospitality during the course of this work, while M.G. 
would like to thank the Aspen Center for Physics, which is supported by National Science Foundation grant PHY-1607611, for hospitality during the later stages of this work. The work of M.~G. was supported by the ERC Starting Grant 679278 Emergent-BH, the Knut and Alice Wallenberg Foundation under grant 113410212 (as Wallenberg Academy Fellow) and the Swedish Research Council grant number 2015-05333.
The work of F.~L.-.M. and K.~Z.  was supported by the grant ``Exact Results in Gauge and String Theories'' from the Knut and Alice Wallenberg foundation. The work of K.~Z. was supported by the ERC advanced grant No 341222, by the Swedish Research Council (VR) grant
2013-4329 and by RFBR grant 15-01-99504. 

\appendix
\renewcommand{\theequation}{\Alph{section}.\arabic{equation}}

\section{The non-relativistic conformal group \label{schgp}}
\label{sec:appgr}

As explained in the main text, we can define the Schr\"{o}dinger group by starting from the relativistic conformal group $SO(d,2)$. The latter
has generators $P_\mu, K_\mu, M_{\mu\nu}, D$, with commutation relations 

\be
[D,P_\mu] =  i P_\mu \;, \;\;\;\;\;\;\; [D,K_\mu] = - i K_\mu \;, \;\;\;\;\;\;\;
[K_\mu,P_\nu] = 2 i (\eta_{\mu\nu} D - M_{\mu\nu})
\ee

\be
[M_{\mu\nu},P_\rho] = -i (\eta_{\mu\rho} P_\nu - \eta_{\nu\rho} P_\mu) \;, \;\;\;\;\; [M_{\mu\nu},K_\rho] = - i (\eta_{\mu\rho} K_\nu - \eta_{\nu\rho} K_\mu) 
\ee

\be
[M_{\mu\nu},M_{\rho\s}] = -i(\eta_{\mu\rho} M_{\nu\s} - \eta_{\nu\rho} M_{\mu\s} + \eta_{\nu\s} M_{\mu\rho} - \eta_{\mu\s} M_{\nu \rho})
\ee

\be
[D,M_{\mu\nu}] = [P_\mu,P_\nu] = [K_\mu, K_\nu] =0 
\ee
Next, we define the lightcone coordinates

\be
x^\pm = \frac{ x^{d-1} \pm x^0}{\sqrt{2}}
\ee
such that the Minkovski metric has components $\eta_{+-}=1, \; \eta_{ij} = \d_{ij}$ with $i, j \in \{1, \ldots, d-2\}$. The non-relativistic conformal or Schr\"{o}dinger group is the subgroup of $SO(d,2)$ that consists of those generators that commute with $N= P^+= P_- = -i \p_-$. 

\be
H = P_+ \;, \;\;\;\;\;\;\; P_i = P_i \;, \;\;\;\;\;\;\; N = P_-
\ee

\be
G_i = M_{i-} \;, \;\;\;\;\;\;\; \mathcal{D} = D + M_{+-}\;, \;\;\;\;\;\;\;M_{ij} = M_{ij}\;, \;\;\;\;\;\;\;C = \half\, K_-
\ee
The non-trivial commutators are  

\be
[\mathcal{D},P_i] = i P_i \;, \;\;\;\;\;\;\; [\mathcal{D},G_i] = -i G_i \;, \;\;\;\;\; [C,P_i] = i G_i
\ee

\be
[\mathcal{D},C] =  -2 i C \;, \;\;\;\;\; [\mathcal{D},H] = 2 i H \;, \;\;\;\;\; [C,H] = i \mathcal{D}
\ee

\be
[M_{ij},P_k] = - i (\d_{ik} P_j - \d_{jk} P_i) \;, \;\;\;\;\;\;\; [M_{ij},G_k] = - i (\d_{ik} G_j - \d_{jk} G_i)
\ee

\be
[M_{ij},M_{kl}] = - i (\d_{ik} M_{jl} - \d_{il} M_{jk} + \d_{jl} M_{ik} - \d_{jk} M_{il})
\ee

\be
[P_i, G_j] =  i \d_{ij} N \;, \;\;\;\;\;\;\; [H,G_i] = - i P_i
\ee
Note that $C, \mathcal{D}$ and $H$ form an $\sl2r$ subalgebra. In terms of the usual generators $L_0, L_{\pm 1}$, we have

\be
C = -i L_1 \;,\;\;\;\;\;H = -i L_{-1} \;,\;\;\;\;\; \mathcal{D} = 2 i L_0
\ee

\section{The $\mathfrak{sl}(2)$ subalgebras}
\label{sec:appsl2}

The non-relativistic conformal group picks out one of the two commuting $\mathfrak{sl}(2)$ subalgebras of $SO(d,2)$, namely
\begin{equation}
 \mathfrak{sl}(2)_+=\left\{ D+ M_{+-},\, P_+, \, K_-\right\}
\end{equation}
The other commuting $\mathfrak{sl}(2)$ subalgebra is given by
\begin{equation}
 \mathfrak{sl}(2)_-=\left\{ D- M_{+-},\, P_-, \, K_+\right\}
\end{equation}
The non-relativistic dilatation generator, whose eigenvalues are the scaling dimensions, is given by $D+ M_{+-}$ from the $\mathfrak{sl}(2)_+$ subalgebra. Its eigenstates will be also chosen to have a definite value of the momentum $P_-$, which is possible as these two subalgebras $\mathfrak{sl}(2)_+$ and $\mathfrak{sl}(2)_-$ commute with each other.

The generators above are related to the conventional generators of $\mathfrak{sl}(2)$ as
\begin{equation}\label{momentum-identif}
 P_\pm = -i L_{-1} = -iS^- ,\qquad K_\mp= -2 i L_1 =-2iS^+,\qquad D\pm M_{+-}= 2 i L_0 = - 2iS^0
\end{equation}
where the commutation relations of $\mathfrak{sl}(2)$ are
\begin{equation}\label{sl2-commrel}
 [S^0,S^\pm]=\pm S^\pm, \qquad [S^+,S^-]=-2S^0
\end{equation}
or, in terms of the $L_i$ generators

\be
[L_1,L_{-1}] = 2 L_0 \;, \;\;\;\;\;\; [L_0,L_{\pm 1}] = \mp L_{\pm 1}
\ee
In the defining two-dimensional representation,
\begin{equation}
 S^0=\frac{\sigma ^3}{2}\,,\qquad S^\pm=\mp\sigma ^\pm.
\end{equation}
The highest-weight unitary representation of $\mathfrak{sl}(2)$ can be constructed with the help of differential operators
\begin{equation}\label{spin-s}
 S^0=z\partial +s,\qquad S^-=\partial ,\qquad S^+=z^2\partial +2sz.
\end{equation}
whose Casimir is

\be
(S^0)^2 - \half (S^+ S^- +S^- S^+) = s(s-1)
\ee
 The discrete series representations of $\mathfrak{sl}(2,\mathbb{R})$ correspond to positive half-integer $s$. The spin chain of the $\mathfrak{sl}(2)$ sector hosts the $s=1/2$ representation at each site. At negative half-integer $s$ the differential operators above realize finite-dimensional unitary representations of $\mathfrak{su}(2)$ on polynomials of degree $2s$.

\section{Scales in the Baxter equation}
\label{sec:scales}

As discussed in the main text, we can fix the Q-function simply by requiring its regularity and solving the Baxter equation perturbatively at small $\dl$. In this approach we never have to deal with the large $u$ asymptotics of the Q-function, however it is instructive to see how the asymptotic behaviour may be linked to our perturbative solution.

Following \cite{Gromov:2016rrp} to solve the Baxter equation we introduce a new variable
\beq
	v\equiv u\;{\dl M}
\eeq
and have to study separately three scales:
\beqa
	\nn&&{\rm scale\;1:\;}u\to \infty\;\;{\rm before}\;\;\dl M\to 0\\
	\nn&&{\rm scale\;2:\;}\dl M\to 0\;\;{\rm with}\;\;v\;\;{\rm fixed}\\
	\nn&&{\rm scale\;3:\;}\dl M\to 0\;\;{\rm then}\;u\to\infty
\eeqa
At scale 1 the solution is captured by the large $u$ asymptotics \eq{Qlarge}. At scale 3 it is $Q=1$ at leading order while at higher orders it is a polynomial as described in section \ref{sec:pertq}.
 
At scale 2 the solution to leading order is captured by the differential equation in $v$
\beq
	v f''(v)+2 f'(v)+2 f(v)=0
\eeq
obtained from the Baxter equation \eq{Baxund} with
\beq
	Q(u)=f(\dl M u)
\eeq
Its solutions are
\beq
	\frac{J_1\left(\sqrt{8v}\right)}{\sqrt{v}},\ \ \ \frac{Y_1\left(\sqrt{8v}\right)}{\sqrt{v}}
\eeq
which are consistent with the asymptotics \eq{Qlarge} at large $v$. Requiring that at small $v$ the solution should reduce to the one at scale 3 which is a polynomial, we see that the true Q-function at scale 2 to leading order is
\beq
	Q(u)={\rm const} \frac{J_1\left(\sqrt{8v}\right)}{\sqrt{v}}\ .
\eeq
It interpolates between the polynomial solution and the complicated large $u$ asymptotics \eq{Qlarge}.

\bibliographystyle{nb}

\end{document}